\documentclass[11pt]{article}

\usepackage{amsmath,amssymb}

\usepackage[dvips]{graphicx}
\newcommand{\f}[2]{\frac{#1}{#2}}
\newcommand{\la}{\langle}
\newcommand{\ra}{\rangle}

\newcommand{\W}{{\cal W}}
\newcommand{\Oc}{{\cal O}}
\newcommand{\de}{\partial}
\newcommand{\tr}{{\rm tr}\,}
\renewcommand{\Re}{{\rm Re}\,}

\newcommand{\I}{\mathbf{1}}
\newcommand{\modbes}{\widetilde{I}}
\newcommand{\barT}{\bar{T}}
\newcommand{\ad}{{\rm Ad}}
\newcommand{\Q}{q}
\newcommand{\qq}{Q}

\newcommand{\zpar}[4]{(#1 #2 | #3 #4)}

\newcommand{\secspace}{\vspace{0.5cm}}

\makeatletter
\@addtoreset{equation}{section}
\makeatother

\begin{document}

\renewcommand{\thefootnote}{\fnsymbol{footnote}}

{\center \LARGE Renormalisation of gauge theories on general anisotropic
  lattices and high-energy scattering in QCD \\} 
{\center \Large Matteo Giordano\footnote{e-mail: giordano@atomki.mta.hu} \\}
{\center \large Institute for Nuclear Research of the
  Hungarian Academy of Sciences \\ 
Bem t\'er 18/c, H-4026 Debrecen, Hungary\\} 
{\center \today \\}
\renewcommand{\thefootnote}{\arabic{footnote}}
\setcounter{footnote}{0}


\begin{abstract}
  We study the renormalisation of $SU(N_c)$ gauge theories on general
  anisotropic lattices, to one-loop order in perturbation theory,
  employing the background field method. The results are then applied
  in the context of two different approaches to hadronic high-energy
  scattering. In the context of the Euclidean nonperturbative approach
  to soft high-energy scattering based on Wilson loops, we refine the
  nonperturbative justification of the analytic continuation relations
  of the relevant Wilson-loop correlators, required to obtain physical
  results. In the context of longitudinally rescaled actions, we
  study the consequences of one-loop corrections on the relation between
  the $SU(N_c)$ gauge theory and its effective description in terms of
  two-dimensional principal chiral models. 
\end{abstract}

\secspace
\section{Introduction}

Anisotropic lattices are a standard tool in modern lattice calculations, 
and have been used in the study of a large variety of problems, 
ranging from glueball~\cite{MP} and light-hadron~\cite{Lin:2008pr}
spectroscopy to properties of QCD 
at finite temperature~\cite{Namekawa:2001ih,Borsanyi:2014vka}.
Numerical calculations in four dimensions usually employ lattices with
3+1 anisotropy,  i.e., only one of the lattice spacings is different
from the others, while more general anisotropy classes have received
much less attention~\cite{Bur}, due to the increasing difficulty in
the scale setting procedure. Indeed, for anisotropy classes other than
3+1, one needs to appropriately tune the action in order to recover
Lorentz invariance in the continuum, already at the pure-gauge theory
level. A better understanding of these more general anisotropy classes
would be useful, since they provide a more flexible setting for
varying length scales independently in different directions. 
This would allow, for example, to enlarge the range of momenta
accessible to lattice calculation at a reasonable computational cost,
by improving the resolution only in a single spatial
direction~\cite{Bur}.  

Anisotropic lattices provide, quite obviously, the natural setting for 
the nonperturbative study of anisotropic systems, also beyond
numerical applications. An interesting case is that of
longitudinally rescaled actions, which in recent years have been
considered in the context of high-energy scattering in
QCD~\cite{Verlinde,Arefeva1,Arefeva2,Orland:2008vg, 
Orland,Cubero:2011ut,Cubero:2012nw}. The basic idea of
Refs.~\cite{Verlinde,Arefeva1,Arefeva2,Orland:2008vg} is to 
perform a rescaling of the longitudinal directions, which appear
highly Lorentz-contracted in a high-energy scattering process, in
order to derive an effective action starting from QCD. In
Refs.~\cite{Verlinde,Arefeva1,Arefeva2,Orland:2008vg} only the
classically rescaled action was considered, while the important effect
of quantum corrections was studied later in
Refs.~\cite{Orland,Cubero:2011ut,Cubero:2012nw}, in the framework of
Wilsonian anisotropic renormalisation in the continuum. In this
context, the use of a gauge-invariant, anisotropic lattice
regularisation could lead to more insight in the structure of quantum
corrections. The relevant anisotropy class here is 2+2, with different
lattice spacings in the longitudinal and in the transverse plane. This
is also the case considered in Ref.~\cite{Bur}, although for different
purposes. 
 
In Ref.~\cite{GM2009} a classical anisotropic rescaling of the
functional integral has been used to justify, on nonperturbative
grounds, the analytic continuation from Euclidean to Minkowski
space, required to obtain physical results in the Euclidean
formulation~\cite{analytic1,analytic2,analytic3,Meggiolaro05,crossing1,crossing2} 
of the nonperturbative approach to soft high-energy
scattering~\cite{Nachtmann91,DFK,Nachtmann97,BN,Dosch,LLCM1,pomeron-book,
reggeon}. This approach has been recently used in Ref.~\cite{sigtot}
to obtain a theoretical estimate of the leading energy dependence 
of hadronic total cross sections, resulting in fair agreement with
experiments. As the analytic continuation plays a key role in this
approach, it is important to establish its correctness going beyond
the formal argument of Ref.~\cite{GM2009}, which, as we have said
above, is based only on a classical rescaling of the QCD action. To
this end, quantum corrections to the effective action must be
included to prove that the necessary analyticity requirements are actually
fulfilled. The relevant anisotropy class in this case is 2+1+1, with
different lattice spacings in the transverse plane and in the two
longitudinal directions.

The purpose of this paper is to perform the renormalisation of a
$SU(N_c)$ gauge theory regularised on a general anisotropic lattice,
and to apply the results in the study of hadronic high-energy
scattering through the approaches mentioned above. To avoid the
complications related to the introduction of fermions on the lattice,
we work here in the {\it quenched} approximation, i.e., pure gauge
theory.  

The plan of the paper is the following. In Section \ref{sec:anis_ren}
we study renormalisation for a general anisotropic lattice
regularisation, using the background field method on the
lattice~\cite{DeW1,DeW2,Abbott1,Abbott2,DG,HH,Karsch,GAKA,LW}. In Section
\ref{sec:AC} we use the results in the context of the nonperturbative
approach to soft high-energy scattering of
Refs.~\cite{Nachtmann91,DFK,Nachtmann97,BN,Dosch,LLCM1,pomeron-book, 
reggeon}, refining the argument of Ref.~\cite{GM2009} on the
possibility of performing analytic continuation to Euclidean space. 
In Section \ref{sec:2dmodel} we discuss the longitudinally rescaled
actions of Refs.~\cite{Verlinde,Arefeva1,Arefeva2,Orland:2008vg,
Orland,Cubero:2011ut,Cubero:2012nw}, focussing on the representation
of the $SU(N_c)$ gauge theory as a set of coupled two-dimensional
principal chiral models. Finally, Section \ref{sec:concl} contains our
conclusions and prospects. Some technical details are discussed in the
Appendices.

\secspace
\section{Anisotropic renormalisation}
\label{sec:anis_ren}

Our aim is to renormalise the Euclidean $SU(N_c)$ gauge theory
regularised on a 4D orthogonal   
an\-iso\-tro\-pic lattice. More precisely, lattice points are located at
$x=\sum_{\mu=1}^4 x_\mu \,\hat\mu$, where $\hat\mu$ are four orthogonal
unit vectors, and the physical coordinates $x_\mu=x_\mu(n)$ in Euclidean space
are $x_\mu(n)=a_\mu n_\mu$, $n_\mu\in \mathbb{Z}$. Here
$a_\mu=a/\lambda_\mu$ is the lattice spacing in direction $\mu$, with
the dimensionless {\it anisotropy parameters} $\lambda_\mu \in \mathbb{R}^+$
being the inverse ratios of $a_\mu$ to a 
reference length scale $a$.\footnote{The five parameters $a$ and
  $\{\lambda_\mu\}$ are obviously redundant, and some condition has to be
  imposed on $\{\lambda_\mu\}$ to remove this redundancy. This notation
  is however convenient, as we treat all the directions on the same
  footing.} 
Consider the following Wilson-like action,  
\begin{equation}
  \label{eq:tree_action}
  S^{\rm tree}_{\rm lat} 
=\beta\sum_n\sum_{\mu<\nu} C_{\mu\nu}
  \left(1 - \f{1}{N_c}\Re\tr U_{\mu\nu}(n)\right)
= \beta\sum_n\sum_{\mu<\nu} C_{\mu\nu}
{\cal P}_{\mu\nu}(n)\,,
\end{equation}
where $U_{\mu\nu}(n)$ are the usual plaquette variables built up 
with the link
variables $U_\mu(n)\in SU(N_c)$,
\begin{equation}
  \label{eq:plaquette}
U_{\mu\nu}(n)=U_\mu(n)U_\nu(n+\hat\mu)U_\mu^\dag(n+\hat\nu)U_\nu^\dag(n)\,,  
\end{equation}
$\beta=2N_c/g^2$ with $g$ the coupling constant, and $C_{\mu\nu}$,
$\mu\neq\nu$, are the {\it plaquette coefficients},\footnote{For
  definiteness, we define also $C_{\mu\mu}=0$.}
\begin{equation}
\label{eq:Ccoeff}
  C_{\mu\nu}=C_{\nu\mu}=C_{\mu\nu}(\lambda) =
  (\lambda_\mu\lambda_\nu)^2\, {\cal J}\,, 
  \qquad {\cal J}^{-1}\equiv \prod_{\alpha=1}^4 \lambda_\alpha\,.
\end{equation}
It is straightforward to show that Eq.~\eqref{eq:tree_action} 
yields the correct na\"ive continuum limit upon identification of the
continuum, physical gauge fields $A_{\mu}(x)$ through
\begin{equation}
\label{eq:links_a}
U_\mu(n) =  e^{iga_\mu A_{\mu}(x(n))}\,, 
\end{equation}
as appropriate for an anisotropic lattice. The choice of plaquette
coefficients $C_{\mu\nu}$ is easily understood by noticing that ${\cal
  J}a^4$ is just the volume of an elementary cell, so that ${\cal J}$
is the Jacobian for the change of variables from isotropic to
anisotropic coordinates, while $a_\mu
a_\nu=(\lambda_\mu\lambda_\nu)^{-1}a^2$ is the area of the faces of an
elementary cell lying in the $\mu\nu$ plane. 

As is well known, divergencies appear in the continuum limit
when taking into account quantum corrections. These divergencies need
to be subtracted through a suitable renormalisation of the couplings
in order to obtain a finite continuum theory. On the isotropic
lattice, the symmetry under the unbroken hypercubic subgroup of $O(4)$
guarantees that all the plaquette terms in the action need to be
renormalised in the same way, so that a single redefinition of $g$ is
sufficient to reabsorb the divergencies. The form of the action is
therefore unchanged, and one recovers the full $O(4)$ invariance in
the continuum limit.  

On a general anisotropic lattice this residual symmetry is
broken, except for reflections through lattice hyperplanes, and
so in general different terms will require a different
renormalisation. Since there are six different plaquette terms and
only four lattice spacings, it will not be possible in the general
case to reabsorb completely the quantum corrections into a
redefinition of $\lambda_\mu$, keeping at the same time the same form
of the tree-level action~\cite{Bur}. In turn, this implies that the
continuum limit of Eq.~\eqref{eq:tree_action} cannot be made into an
$O(4)$-invariant theory by an appropriate, simple rescaling of the
lattice spacings, since in the general case one will still find
different coefficients for the six continuum field-strength terms. To
recover $O(4)$ invariance one must ensure that these coefficients are
equal, and this requires that we take the action to be of the more
general form
\begin{equation}
  \label{eq:action}
  S_{\rm lat} = \sum_n\sum_{\mu<\nu}\beta_{\mu\nu} C_{\mu\nu} 
  \left(1 - \f{1}{N_c}\Re\tr U_{\mu\nu}(n)\right)
= \sum_n\sum_{\mu<\nu}\beta_{\mu\nu} C_{\mu\nu} 
{\cal P}_{\mu\nu}(n)\,,  
\end{equation}
where the couplings
$\beta_{\mu\nu}=\beta_{\nu\mu}=\beta_{\mu\nu}(\lambda)$ have to 
be properly tuned to yield a finite, $O(4)$-invariant theory in the
continuum limit. 

The need for tuning comes, as we have said, from the fact that there are
in general more couplings than anisotropy parameters. It is however
easy to show that one has to tune at most only two combinations of the
couplings to achieve restoration of $O(4)$ invariance in the continuum
limit, while the other four independent combinations can be
interpreted as the coupling fixing the overall lattice scale, and
renormalisations of the anisotropies
$a_\nu/a_\mu=\lambda_\mu/\lambda_\nu$.  
To see this, let us remove the redundancy in the
set $\{\lambda_\mu\}$ by imposing the symmetric condition $\prod_\mu
\lambda_\mu =1$, thus defining $a$ in terms of the volume of an
elementary cell. Any other equivalent choice (i.e., giving the same
$a_\mu$) is obtained by a simple global rescaling of
$\{\lambda_\mu\}$ and of $a$. The six plaquette terms can be grouped
in pairs of ``complementary'' 
${\mu}{\nu}$ and ${\bar\mu}{\bar\nu}$ plaquettes, i.e.,
$(U_{12},U_{34})$, etc., which we denote as 
$\zpar{\mu}{\nu}{\bar\mu}{\bar\nu}=\zpar{1}{2}{3}{4},
\zpar{1}{3}{2}{4},\zpar{1}{4}{2}{3}$. It is also easily noticed that
$C_{\mu\nu}=\f{\lambda_\mu\lambda_\nu}{\lambda_{\bar\mu}\lambda_{\bar\nu}}$,
so that $C_{\bar\mu\bar\nu}=C_{\mu\nu}^{-1}$. This suggests to
parameterise $\beta_{\mu\nu}$ as follows,
\begin{equation}
  \label{eq:zparamet}
  \beta_{\mu\nu} = \beta \, Z_{\zpar{\mu}{\nu}{\bar\mu}{\bar\nu}} \,
  \f{z_\mu z_\nu}{z_{\bar\mu} z_{\bar\nu}}\,.
\end{equation}
As there are two redundant parameters, we choose to fix $\prod_\mu
z_\mu =1$, so that our condition on  $\{\lambda_\mu\}$ is not
renormalised,\footnote{Any other
  choice is of course allowed. If, for example, the scale $a$ is defined
  to be one of the lattice spacings by choosing $\lambda_\mu=1$ for
  some $\mu$, then it is convenient to choose $z_\mu=1$. The new
  values of $z_\nu$ are obtained from those corresponding to the
  symmetric condition by replacing $z_\nu\to z_\nu/z_\mu$,
  while $Z_{\zpar{\mu}{\nu}{\bar\mu}{\bar\nu}}$ and $\beta$ are
  unaffected.} and $\prod_{\zpar{\mu}{\nu}{\bar\mu}{\bar\nu}} 
Z_{\zpar{\mu}{\nu}{\bar\mu}{\bar\nu}} =1$. In this way $\beta$, 
$z_\mu$ and $Z_{\zpar{\mu}{\nu}{\bar\mu}{\bar\nu}}$ are unambiguously
defined and can be obtained from $\beta_{\mu\nu}$ as follows,
\begin{equation}
  \label{eq:zbeta}
  \beta =
  \left(\prod_{\mu<\nu}\beta_{\mu\nu}\right)^{\f{1}{6}}\,,\quad
  z_\mu =
  \left(\prod_{\nu\neq\mu}\f{\beta_{\mu\nu}}{\beta_{\bar\mu\bar\nu}}\right)^{\f{1}{8}}\,,
 \quad  
 Z_{\zpar{\mu}{\nu}{\bar\mu}{\bar\nu}} = \left[
\f{\beta_{\mu\nu}\beta_{\bar\mu\bar\nu}}{\left(
\beta_{\mu\bar\nu}\beta_{\bar\mu\nu}\beta_{\mu\bar\mu}\beta_{\nu\bar\nu}
\right)^{\f{1}{2}}}
\right]^{\f{1}{3}}\,.
\end{equation}
This makes it explicit that the restoration or not of $O(4)$
invariance in the continuum depends only on the values of the ratios
of the couplings $\beta_{\mu\nu}$. Defining now the bare
anisotropy parameters $\lambda_\mu^B \equiv z_\mu\lambda_\mu$, and the bare
plaquette coefficients 
$C_{\mu\nu}^B  \equiv C_{\mu\nu}(\lambda^B)$, one can
rewrite  Eq.~\eqref{eq:action} as
\begin{equation}
  \label{eq:zparamet2}
  S_{\rm lat} = \beta\!\!
  \sum_{n,\zpar{\mu}{\nu}{\bar\mu}{\bar\nu}}Z_{\zpar{\mu}{\nu}{\bar\mu}{\bar\nu}}  
  \left[C^B_{\mu\nu}P_{\mu\nu}(n) + C^B_{\bar\mu\bar\nu}P_{\bar\mu\bar\nu}(n)\right]\,.
\end{equation}
This equation shows that to obtain an $O(4)$-invariant theory in the
continuum limit, one can choose freely $\lambda_\mu^B$ (up to
a constraint to remove the redundancy), and then tune only the two
independent ratios of $Z_{\zpar{\mu}{\nu}{\bar\mu}{\bar\nu}}$ to the
appropriate values. The physical anisotropy parameters $\lambda_\mu$ are
related to the bare ones through the renormalisation
$\lambda_\mu=z_\mu^{-1}\lambda_\mu^B$, and can be measured {\it ex post}. 

Using the parameterisation Eq.~\eqref{eq:zbeta}, it is possible to set
up a rather simple nonperturbative scheme to achieve restoration of
$O(4)$ invariance in the continuum, for an arbitrary choice of bare
anisotropy parameters. The basic idea is to impose that the
string tension, determined from the asymptotic behaviour of large
rectangular $T\times R$ on-axis Wilson loops $W_{\alpha\beta}\sim
\exp\{-\hat\sigma_{\alpha\beta}T R\}$, is the same for all pairs of
directions $\alpha,\beta$. Denoting with $\sigma$ the physical
(dimensionful) string tension, this amounts to impose
$\lambda_\alpha\lambda_\beta\hat\sigma_{\alpha\beta} = a^2\sigma$,
for all pairs of different $\alpha,\beta$. Multiplying the relations
for $\hat\sigma_{\alpha\beta}$ and its ``complementary''
$\hat\sigma_{\bar\alpha\bar\beta}$, one obtains the following consistency
conditions, 
\begin{equation}
  \label{eq:consistency_cond}
  \hat\sigma_{12}  \hat\sigma_{34}=   \hat\sigma_{13}
  \hat\sigma_{24}=   \hat\sigma_{14}  \hat\sigma_{23}\,,
\end{equation}
which have to be imposed to recover $O(4)$ invariance. This can be
done without any prior knowledge of the physical $\lambda_\alpha$, and
requires only to properly tune two of the coefficients
$Z_{\zpar{\mu}{\nu}{\bar\mu}{\bar\nu}}$ (the third one being
constrained by our choice $\prod_{\zpar{\mu}{\nu}{\bar\mu}{\bar\nu}} 
Z_{\zpar{\mu}{\nu}{\bar\mu}{\bar\nu}} =1$). Having done this, the
anisotropies can then be obtained from the ratio
$\lambda_\mu/\lambda_\nu=
\hat\sigma_{\nu\alpha}/\hat\sigma_{\mu\alpha}$ for any
$\alpha\neq\mu,\nu$. Imposing $\prod_\mu\lambda_\mu =1$ one can 
explicitly determine all $\lambda_\mu$'s, and set the lattice scale
$a$ from the relation $a^4\sigma^2 = \hat\sigma_{12}
\hat\sigma_{34}$. While the string tension is known not to be the
best observable for setting the physical scale, nevertheless it
could be useful for the tuning, as it can be determined to high
precision by means of multilevel algorithms~\cite{LWml}. 
It is worth mentioning that the tuning of two parameters is only
required when all the lattice spacings are different: if at least a
pair of lattice spacings are equal, one easily sees that only one 
parameter has to be tuned.\footnote{ 
In the 3+1 case, where a single lattice spacing differs from the
others, there are only two kinds of plaquette terms and so only two
independent lattice string tensions. In this case there is thus no
consistency condition to be satisfied and no tuning is needed, as
is well known.}

\subsection{Background field method}

From the discussion above, we see that our task is to find the
relations among the couplings $\beta_{\mu\nu}$ that will lead to an
$O(4)$-invariant theory in the continuum limit. We will study this
problem to lowest order in perturbation theory, making use of the
background field method~\cite{DeW1,DeW2,Abbott1,Abbott2}
on the lattice~\cite{DG,HH,Karsch,GAKA,LW}. The advantage of this
method is that it allows to keep an exact gauge invariance on the
lattice after gauge fixing, which greatly simplifies the
calculations. A full account on the background field method can be 
found elsewhere~\cite{BGF-rev,Z-J}. Here we briefly recall the main
points of the method to fix the notation.  

The first step is to introduce a background field $U^{(c)}_{\mu}$ in
the gauge action as follows, 
\begin{equation}
  \label{eq:bgf_0}
 S_{\rm BF}[U^{(c)},V] \equiv S_{\rm lat}[VU^{(c)}] \,,
\end{equation}
where we now denote with $V$ the gauge links, to be 
integrated over with the usual Haar measure. As a consequence of the
gauge invariance of $S_{\rm lat}$, the action $S_{\rm BF}$ is
invariant under the {\it background gauge transformation}  
\begin{equation}
  \label{eq:BGT}
  U^{(c)\,
G}_{\mu}(n) = G(n)U^{(c)}_{\mu}(n)G^\dag(n+\hat\mu)\,,
  \qquad
  V^G_\mu(n) = G(n)V_\mu(n)G^\dag(n)\,,
\end{equation}
with $G(n)\in SU(N_c)$, as well as under the following gauge
transformation of $V$ alone, 
\begin{equation}
  \label{eq:GT}
    V_\mu(n) \to G(n)V_\mu(n)U^{(c)}_{\mu}(n)
    G^\dag(n+\hat\mu)U^{(c)}_{\mu}{}^\dag(n)\,. 
\end{equation}
The integration measure is also obviously invariant
under the transformations Eqs.~\eqref{eq:BGT} and \eqref{eq:GT}. One
then proceeds to set up perturbation theory in the usual way, setting
\begin{equation}
  \label{eq:parameterisation}
  V_\mu(n) = e^{i\f{g}{\lambda_\mu} \Q_\mu(n)}\,, \qquad
  U^{(c)}_{\,\mu}(n) = e^{ia_\mu B_\mu(n)}\,, 
\end{equation}
where\footnote{Here and in the following, the
  sum over repeated colour indices is understood.} 
$\Q_\mu(n)=\Q^a_\mu(n)t^a$ and $B_\mu(n)=B^a_\mu(n)t^a$, with
$t^a$ the generators of $SU(N_c)$ in the fundamental representation,
$a=1,\ldots, N_c^2-1$, and $\tr\{ t^a t^b\} =
\f{1}{2}\delta^{ab}$. One then changes variables of integration to
$\Q$, expressing the Jacobian as a contribution $S_{\rm meas}[\Q]$ to
the action. Notice that powers of $g$ and $a$ are chosen so that $\Q$
is dimensionless, while $B$ has dimensions of mass. This distinction
is convenient for book-keeping purposes~\cite{GAKA}.  

Under the transformation Eq.~\eqref{eq:BGT}, the
background field $B$ transforms as a gauge field, while the
``quantum'' field $\Q$ transforms as a matter field in the adjoint
representation. The symmetry under the gauge transformation
Eq.~\eqref{eq:GT} requires to impose a gauge condition on $\Q$ to
define the corresponding propagator. This is done {\it \`a la}
Faddeev--Popov, adding a gauge-fixing term to the action, together
with the corresponding ghost term. The key point is that there is an
appropriate choice of gauge, called the {\it background field gauge},
for which the gauge-fixing and the ghost terms are invariant under the
background gauge transformation Eq.~\eqref{eq:BGT}. This gauge-fixing
term is~\cite{DeW1,DeW2,LW} 
\begin{equation}
 \label{eq:gauge_fixing}
  S_{\rm g.f.}[B,\Q] = {\cal J}\sum_n \tr \bigg(\sum_\mu D^-_\mu \Q_\mu\bigg)^2\,,
\end{equation}
where $D^\pm_\mu$ are the lattice background covariant differences,
\begin{equation}
\begin{aligned}
  \label{eq:cov_der}
    D^+_\mu {f}(n) &\equiv
  \lambda_\mu\bigg[U^{(c)}_\mu(n){f}(n+\hat{\mu})U^{(c)}_\mu{}^\dag(n)-
  {f}(n)\bigg],\\ 
  D^-_\mu {f}(n) &\equiv
  \lambda_\mu
  \bigg[U^{(c)}_\mu{}^\dag(n-\hat{\mu}){f}(n-\hat{\mu})
  U^{(c)}_\mu(n-\hat{\mu})-{f}(n)\bigg]\,,
\end{aligned}
\end{equation}
in which a factor $\lambda_\mu$ is also included for convenience. The
usual lattice differences $\Delta^{\pm}_\mu$ are obtained setting
$U^{(c)}_\mu=\I$ in the expressions above, where $\I$ denotes the unit
matrix. The corresponding ghost term is
\begin{equation}
  \label{eq:ghost_full}
S_{\rm ghost}[B,\Q,c,\bar c] =  2 {\cal J}\sum_{n,\mu} \tr
\left\{[D^+_\mu \bar{c}(n)] 
\left[ M^{-1}\left({\textstyle\f{g}{\lambda_\mu}}\Q_\mu(n)\right)D^+_\mu + i\ad
\left({\textstyle\f{g}{\lambda_\mu}}\Q_\mu(n)\right)\right]c(n)\right\}  \,,
\end{equation}
where $c = c^{a} t^a$, $\bar{c} = \bar{c}^{a} t^a$, with $c^a,\bar c^{a}$
independent Grassmann variables, and where
\begin{equation}
  \label{eq:Mmatrix}
  M(X) \equiv \f{1-e^{-i\ad(X)}}{i\ad(X)}\,, \qquad \ad(X)Y \equiv [X,Y]\,.
\end{equation}
It is straightforward to prove invariance of these two terms under the
background gauge transformation, Eq.~\eqref{eq:BGT}, supplemented by
the transformation laws for the ghost fields, 
\begin{equation}
  \label{eq:gh_tr}
c^G(n) =
G(n)c(n)G^\dag(n)\,, \qquad \bar c^{\,G}(n) = G(n)\bar c(n)G^\dag(n)\,.  
\end{equation}
Expanding Eq.~\eqref{eq:ghost_full} up to  $\Oc(g^0)$, one finds 
\begin{equation}
  \label{eq:ghost}
  \begin{aligned}
    S_{\rm ghost}[B,\Q,c,\bar c] &= 2 {\cal J}\sum_{n,\mu} \tr \big\{[D^+_\mu
    \bar{c}(n)][D^+_\mu c(n)]\big\} 
 +\Oc(g) \\ &
= 2{\cal J}\sum_{n,\mu}\tr \big\{\bar{c}(n)\, D^-_{\mu}D^+_\mu c(n)\big\}
 + \Oc(g) \equiv S^0_{\rm ghost}[B,c,\bar c]  + \Oc(g), 
  \end{aligned}
\end{equation}
where we have used ``integration by parts'' on a lattice (infinite or
with periodic boundary conditions), 
\begin{equation}
\label{eq:intbyparts}
    \sum_n \tr \big\{[D^+_\mu f(n)]g(n)\big\} =   \sum_n \tr \big\{ f(n)[D^-_\mu
    g(n)]\big\}\,.
\end{equation}
The starting point for the perturbative analysis is the generating
functional
\begin{equation}
  \label{eq:ZBF}
  \begin{aligned}
       Z[B,J,\bar\eta,\eta] &= \int {\cal D} \Q {\cal D}c {\cal D}\bar
       c\, e^{-S_{\rm tot}[B,\Q,c,\bar c] + J\cdot \Q +\bar\eta \cdot c + 
     \eta\cdot\bar c} = e^{W[B,J,\bar\eta,\eta]}\,,  \\
  S_{\rm tot}[B,\Q,c,\bar c]&= S_{\rm BF}[B,\Q]+S_{\rm meas}[\Q]+
  S_{\rm g.f.}[B,\Q] +S_{\rm ghost}[B,\Q,c,\bar c]\,,
   \end{aligned}
\end{equation}
where with a small abuse of notation we have written $S_{\rm
  BF}[B,\Q]=S_{\rm BF}[U_c,V]$, and we have added source terms for the
various fields. Here $J=J(n)=J^a_\mu(n)t^a$ and $J\cdot \Q \equiv
\sum_{n,\mu} J^a_\mu(n) \Q^a_\mu(n)$, and similarly for the other terms.  
A Legendre transform gives the effective action (generating functional
for 1PI graphs),
\begin{equation}
  \label{eq:effact0}
  \Gamma[B,\qq,C,\bar C] = - W[B,J,\bar\eta,\eta] + J\cdot \qq + \bar\eta
  \cdot C + \bar C \cdot\eta\,, 
\end{equation}
where the classical fields $\qq$, $C$ and $\bar C$ are defined as
\begin{equation}
\qq^a_\mu(n) = \f{\de
    W[B,J,\bar\eta,\eta]}{\de J^a_\mu(n)}\,,\quad
    C^a(n) = \f{{\de} 
    W[B,J,\bar\eta,\eta]}{\de \bar\eta^a(n)}\,,\quad \bar C^a(n) = 
  \f{{\de} 
    W[B,J,\bar\eta,\eta]}{\de \eta^a(n)}\,,
\end{equation}
i.e., they are the expectation values of the quantum fields for
prescribed values of $B$ and of the sources. 

Defining a background gauge transformation for the classical fields,
imposing that they transform as the corresponding quantum fields,
Eqs.~\eqref{eq:BGT} and \eqref{eq:gh_tr}, leads
finally to the identity
\begin{equation}
  \label{eq:Ginv}
  \Gamma[B^G,\qq^G,C^G,\bar C^G]=\Gamma[B,\qq,C,\bar C]
\end{equation}
for the effective action. This is the key relation that allows us to
simplify the calculations. Indeed, setting $S_{\rm
  eff}[B]\equiv\Gamma[B,0,0,0]-\Gamma[0,0,0,0]$, as a consequence of
the background gauge invariance, of the discrete symmetries of the
action (translations and reflections\footnote{Reflections
  $\Pi_\alpha$ act as follows on the coordinates, $\Pi_\alpha
  n_\mu = n_\mu$ for $\mu\neq\alpha$, $\Pi_\alpha
  n_\alpha = -n_\alpha$. The corresponding transformation laws for $B$
  and $\Q$ are the following, 
 \begin{equation}
    \label{eq:reflections}
    B^{\Pi_\alpha}_\mu(n) = \left\{
       \begin{aligned}
        & B_\mu(\Pi_\alpha n)\quad \mu\neq\alpha\,,\\
        & -B_\alpha(\Pi_\alpha n -\hat \alpha)\,, 
      \end{aligned}
    \right.
    \qquad
    \Q^{\Pi_\alpha}_\mu(n) = \left\{
      \begin{aligned}
        & \Q_\mu(\Pi_\alpha n)\quad \mu\neq\alpha\,,\\
        & -U^{(c)}_\alpha{}^\dag(\Pi_\alpha n -\hat \alpha)
        \Q_\alpha(\Pi_\alpha n -\hat \alpha)U^{(c)}_\alpha(\Pi_\alpha n -\hat
        \alpha)\,. 
      \end{aligned}\right. 
   \end{equation}
}), and of the locality of
divergencies, to one-loop accuracy and to lowest order in perturbation
theory we are guaranteed to find in the continuum limit
\begin{equation}
  \label{eq:bfm_main}
  \begin{aligned}
    \lim_{a\to 0} S_{\rm  eff}[B]  =&~ \f{1}{2}\sum_{\mu,\nu} \int d^4x
    \left[\f{\beta_{\mu\nu}}{2N_c}
    - { K}_{\mu\nu}\right]
    \tr {\cal F}^{\,2}_{\mu\nu}(x)
    \\ & \phantom{auauauau}
+ \text{(non-local finite terms)} + \Oc(g^2)
    \,, 
  \end{aligned}
\end{equation}
where ${K}_{\mu\nu}={K}_{\nu\mu}={K}_{\mu\nu}(a,\lambda)$ is $\Oc(g^0)$, and
where ${\cal F}_{\mu\nu}= \de_\mu {\cal B}_\nu - \de_\nu {\cal B}_\mu
+ i[{\cal B}_\mu,{\cal B}_\nu]$ is the field strength for the
continuum background field ${\cal B}_\mu(x)$, ${\cal B}_\mu(x(n)) =
B(n)$. 
For our purposes it is therefore sufficient to compute the 
two-point function of the background field to have enough
information to renormalise the theory and impose $O(4)$ invariance. To
one-loop accuracy it is enough to set
\begin{equation}
  \label{eq:restore1}
  \f{\beta_{\mu\nu}}{2N_c} - { K}_{\mu\nu} = \f{1}{g^2} +
\f{\delta\beta_{\mu\nu}}{2N_c}  - {K}_{\mu\nu} = \f{1}{g_r^2}\,,
\end{equation}
where $g_r$ is the renormalised, $\lambda$-independent coupling. 

We notice that Eq.~\eqref{eq:action}, with the couplings chosen
according to Eq.~\eqref{eq:restore1}, can be interpreted in two
ways. Under the identification $U_\mu(n) =  e^{iga_\mu
  A_{\mu}(x(n))}$ with $x_\alpha=a_\alpha n_\alpha$, it leads in the
continuum to the renormalised, {\it isotropic} action for the gauge
fields $A_{\mu}(x)$, for which it provides an appropriate lattice
discretisation. On the other hand, identifying $U_\mu(n) = e^{iga
  \phi_{\mu}(y(n))}$ with $y_\alpha =a n_\alpha$, in the continuum
limit one obtains the following renormalised {\it anisotropic} action, 
\begin{equation}
  \label{eq:cont_ani}
S \to   \f{1}{2g^2_r}\sum_{\mu,\nu} C_{\mu\nu}\int d^4y\,
        \tr { \Phi}^{\,2}_{\mu\nu}(y)
\,,
\end{equation}
with ${\Phi}_{\mu\nu}$ the usual field-strength tensor for
$\phi_\mu$, for which Eq.~\eqref{eq:action} provides therefore a lattice
discretisation. This is the form of the action obtained by classically
rescaling coordinates and fields in the Yang-Mills action, discussed
in Refs.~\cite{Verlinde,Arefeva1,Arefeva2,Orland:2008vg,GM2009}.

\subsection{One-loop calculation}

To compute ${K}_{\mu\nu}$ it is enough to expand the action to
order $\Oc(g^0)$, which in turn means expanding the gauge action up to 
second order in $\Q$. Contributions from $S_{\rm meas}$ are at least
$\Oc(g^2)$ and can be ignored. Let us expand the action $S_{\rm BF}+S_{\rm
  g.f.}$ in powers of $\Q$,
\begin{equation}
  \label{eq:act_expansion}
S_{\rm BF}[B,\Q]+S_{\rm g.f.}[B,\Q] = S_c[B] +
S_{{\rm g}1}[B,\Q] + S_{{\rm g}2}[B,\Q]+\ldots\,,  
\end{equation}
 where 
$S_c[B]=S_{\rm BF}[B,0]$ is the classical action, 
$S_{{\rm g}1}[B,\Q]$
is linear in $\Q$, $S_{{\rm g}2}[B,\Q]$ is quadratic and so on, and set
\begin{equation}
  \label{eq:quad_part}
  \begin{aligned}
    S_2[B,\Q,c,\bar c] &= S_{{\rm g}2}[B,\Q] + S^0_{\rm
      ghost}[B,c,\bar c] \\
 &= \sum_{n, m, \mu, \nu} \f{1}{2}
    \Q_\mu^a(n)\big(\Pi[B]\big)^{ab}_{nm;\mu\nu}\Q_\nu^b(m)
+  \sum_{n, m}
    \bar c^a(n) \big(\hat\Pi[B]\big)^{ab}_{nm}  c^b(m)\,.
  \end{aligned}
\end{equation}
A straightforward calculation then
shows that 
\begin{equation}
  \label{eq:G_BF1l2}
  \begin{aligned}
    S_{\rm eff}[B]\big|_{\Oc(g^0)} &=S_{\rm BF}[B,0]  +\f{1}{2}\log
    \f{\det \Pi[B]}{\det \Pi[0]} - 
    \log\f{\det \hat\Pi[B]}{\det \hat\Pi[0]} \,. 
  \end{aligned}
\end{equation}
Terms linear in $\Q$ play no role and can be ignored.\footnote{These
  terms are usually discarded by requiring $B$ to satisfy the
  equations of motion, but this is actually not necessary.}
Eq.~\eqref{eq:G_BF1l2} can be conveniently written as 
\begin{equation}
  \label{eq:G_BF1l3}
    e^{-S_{\rm eff}[B]}\big|_{\Oc(g^0)} =
    e^{-S_c[B]}  \la e^{- (S_2 - S^{\rm free})  }\ra_0
    \,,
\end{equation}
where $S^{\rm free}[\Q,c,\bar c]= S_2[0,\Q,c,\bar c] $ is the free action 
with no background field, and $\la \ldots\ra_0$ denotes the
corresponding expectation value,
\begin{equation}
  \label{eq:ave}
  \begin{aligned}
  \la{\cal O}[B,\Q,c,\bar c]\ra_0 &= Z_{\rm free}^{-1}\int {\cal D} \Q
  {\cal D}c {\cal D}\bar c\, 
e^{-S^{\rm free}[\Q,c,\bar c]}\, {\cal O}[B,\Q,c,\bar c]\,,\\
Z_{\rm free}&=\int {\cal D}\Q {\cal D}c {\cal D}\bar
       c\,  e^{-S^{\rm free}[\Q,c,\bar c]}\,.
  \end{aligned}
\end{equation}
For future utility, we define the connected correlation function $\la
\Oc_1 \Oc_2\ra_{0\,c} \equiv \la \Oc_1 \Oc_2\ra_0 - \la \Oc_1
\ra\la\Oc_2\ra_0$. 
Since we are interested only in the two-point function for $B$, only
terms up to $\Oc(B^2)$ will be kept in $S_2$.

\subsubsection{The quadratic action}

The gauge action in a background field can be conveniently written as
follows, 
\begin{equation}
S_{\rm BF}[B,\Q]=  S_{\rm lat}[VU^{(c)}] = \sum_{n,\mu<\nu} \beta_{\mu\nu} C_{\mu\nu} 
  \left(1-\frac{1}{N_c}\Re\tr \big\{V_{\mu\nu}(n)U^{(c)}_{\mu\nu}(n)\big\}\right),
\end{equation}
where the ``quantum'' and the ``background'' plaquette are given
respectively by
\begin{equation}
\begin{aligned}
  V_{\mu\nu}(n) &\equiv e^{-ig\frac{1}{\lambda_\mu}\left(\f{1}{\lambda_\nu}D^+_\nu
    \Q_\mu(n) + \Q_\mu(n)\right)} e^{-ig\frac{1}{\lambda_\nu}\Q_\nu(n)} 
  e^{ig\frac{1}{\lambda_\mu}\Q_\mu(n)}
  e^{ig\frac{1}{\lambda_\nu}\left(\f{1}{\lambda_\mu} D^+_\mu \Q_\nu(n) + 
    \Q_\nu(n)\right)},\\ 
  U^{(c)}_{\mu\nu}(n) &\equiv
  U^{(c)}_{\mu}(n)U^{(c)}_{\nu}(n+\hat{\mu})
  U^{(c)}_{\mu}{}^\dag(n+\hat{\nu})U^{(c)}_{\nu}{}^\dag(n)\,.  
  \phantom{\bigg[\bigg]} 
\end{aligned}
\end{equation}
A standard application of the Baker-Campbell-Hausdorff formula gives 
\begin{equation}
  \label{eq:bgplaq}
  U^{(c)}_{\mu\nu}(n) =
  \exp\left\{i\frac{a^2}{\lambda_\mu\lambda_\nu}f_{\mu\nu}(n) + 
    {\cal O}(a^3B\de B,a^4(\de B)^2) + \Oc(a^3B^3)
  \right\},
\end{equation}
with\footnote{In the following equations we will sometimes drop the
  dependence on the lattice site $n$ to make the expressions more
  readable.} 
\begin{equation}
  f_{\mu\nu} = a^{-1}\left(\Delta^+_\mu B_\nu - \Delta^+_\nu
    B_\mu\right) + i[B_\mu,B_\nu],
\end{equation}
which in the continuum limit reduces to the usual f\mbox{}ield strength
tensor for the background f\mbox{}ield.\footnote{
  \label{foot:ho0} 
  In principle, also the higher-order terms of order ${\cal O}(a^3B\de
  B,a^4(\de B)^2)$ appearing in Eq.~\eqref{eq:bgplaq} could 
  contribute to the two-point function in the continuum. This
  however is not the case, as we will see below (see footnotes
  \ref{foot:ho1} and \ref{foot:ho2}).} For $V_{\mu\nu}$ we have
instead   
\begin{equation} 
\label{eq:Vmunu}
 V_{\mu\nu}(n) =
 \exp\left\{ig\frac{1}{\lambda_\mu\lambda_\nu}\left[F_{\mu\nu}(n) +
   gR_{\mu\nu}(n)\right] + 
  {\cal O}(g^3)\right\},
\end{equation}
where
\begin{equation}
\label{eq:FRdef}
\begin{aligned}
  F_{\mu\nu} &= D^+_\mu \Q_\nu - D^+_\nu \Q_\mu,\\
  R_{\mu\nu}^{(1)} &= \frac{i}{2\lambda_\mu\lambda_\nu}[D^+_\mu
    \Q_\nu, D^+_\nu \Q_\mu] + i[\Q_\mu, \Q_\nu],\\
  R_{\mu\nu}^{(2)} &= \frac{i}{2} \left(\frac{1}{\lambda_\mu}[\Q_\mu,D^+_\nu
    \Q_\mu] - \frac{1}{\lambda_\nu} [\Q_\nu, D^+_\mu \Q_\nu]\right)\,,
\end{aligned}
\end{equation}
and $R_{\mu\nu} = R_{\mu\nu}^{(1)} + R_{\mu\nu}^{(2)}$. 
Expanding up to quadratic terms in $B$ and $\Q$ we find
\begin{equation}
S_{\rm BF}[B,\Q] =  S_c[B] + S_{q}[B,\Q]  + \text{(linear in $\Q$)}
+ \Oc(\Q^3)\,,
\end{equation}
where $S_c$ is the classical action, already defined above,
\begin{equation}
  S_c = {\cal J}a^4\sum_{n,\mu,\nu}
  \frac{\beta_{\mu\nu}}{2N_c}\,\frac{1}{2}\tr f^{\,2}_{\mu\nu}(n) \quad \mathop
  \to_{a\to 0} \quad \frac{1}{2}\int d^4x\,\sum_{\mu,\nu}
  \frac{\beta_{\mu\nu}}{2N_c}\,\tr {\cal F}^{\,2}_{\mu\nu}(x)\,,
\end{equation}
while the ``quantum'' piece $S_q$ is given by
\begin{equation}
  S_{q}[B,\Q] = {\cal
    J}\sum_{n,\mu,\nu}\frac{1}{2}\tr\big\{F^{\,2}_{\mu\nu}(n) + 
  2a^2R_{\mu\nu}(n)f_{\mu\nu}(n)\big\} - \frac{1}{4}
  \frac{a^4}{(\lambda_\mu\lambda_\nu)^2}\,\tr \big\{ F_{\mu\nu}^{\,2}(n)
  f_{\mu\nu}^{\,2}(n)\big\}\,.
\end{equation}
The gauge-fixing term is quadratic in $\Q$, and can be conveniently
rearranged as follows,
\begin{equation}
      S_{\rm g.f.} =   S_{\rm g.f.}^{(1)} +   S_{\rm g.f.}^{(2)} +
    S_{\rm T'},
\end{equation}
where 
\begin{equation}
\begin{aligned}
  S_{\rm g.f.}^{(1)} &=   {\cal J}\sum_{n,\mu,\nu} \tr \big\{D^+_\nu \Q_\mu(n) D^+_\mu
  \Q_\nu(n)\big\}\,, \qquad
  S_{\rm g.f.}^{(2)} = {\cal J}a^2 \sum_{n,\mu,\nu}
  \tr\Big\{\bar R^{(1)}_{\mu\nu}(n)f_{\mu\nu}(n)\Big\}\,, 
\\
  \bar R^{(1)}_{\mu\nu} &= i\left([\Q_\mu,\Q_\nu] +
  \frac{1}{\lambda_\mu}[\Q_\mu, D^+_\mu 
      \Q_\nu] 
    - \frac{1}{\lambda_\nu}[\Q_\nu, D^+_\nu \Q_\mu] + 
    \frac{1}{\lambda_\mu\lambda_\nu}[D^+_\nu \Q_\mu, D^+_\mu
      \Q_\nu]\right)\,,\\
S_{\rm T'} &= {\cal J}a^4 \sum_{n,\mu,\nu} \tr \bar
R^{(2)}_{\mu\nu}(n)\\
  \bar R^{(2)}_{\mu\nu} &= 
  \frac{i}{2\lambda_\mu\lambda_\nu}
  [f_{\mu\nu},\frac{1}{\lambda_\mu}D^+_\mu \Q_\nu +
  \Q_\nu] 
  [f_{\mu\nu},\frac{1}{\lambda_\nu}D^+_\nu \Q_\mu +
  \Q_\mu]\,.
\end{aligned}
\end{equation}
Finally, the ghost term is independent of $\Q$ to $\Oc(g^0)$. Putting
all the terms together, one obtains for the quadratic lattice action 
\begin{equation}
    S_2 = S_c + S^{\rm free} +
  S_{\rm gluon}^{\rm int} + S_{\rm ghost}^{\rm int} + S_{\rm A} + S_{\rm B} +
  S_{\rm T} + S_{\rm T'},
\end{equation}
where the terms have been grouped so that each quantity in the
equation above is separately invariant under a background gauge
transformation~\cite{GAKA}. Here $ S^{\rm free}=S_{\rm gluon}^{\rm free} +
S_{\rm ghost}^{\rm free}$, with 
\begin{equation}
\label{eq:free_action}
 S_{\rm gluon}^{\rm free} = 
{\cal J}\sum_{n,\mu,\nu} \tr
  \big[\Delta^+_\mu \Q_\nu(n)\big]^2 
  \,, \qquad 
  S_{\rm ghost}^{\rm free} = 2{\cal J}\sum_{n,\mu,\nu} \tr \big\{\bar{c}(n)\,
  \Delta^-_\mu\Delta^+_\mu c(n)\big\} 
  \,,
\end{equation}
being the free actions for gluons and ghosts, respectively, in
terms of which the propagators are def\mbox{}ined, while the
interaction terms are given by
\begin{equation}
\label{eq:int_action}
\begin{aligned}
  S_{\rm gluon}^{\rm int} &= S_q  + S_{\rm g.f.}^{(1)} 
- S_{\rm gluon}^{\rm free} = {\cal J}\sum_{n,\mu,\nu} 
 \tr \big\{[D^+_\mu \Q_\nu(n)]^2 -   [\Delta^+_\mu \Q_\nu(n)]^2\big\}\,,  \\ 
  S_{\rm ghost}^{\rm int} &= S^0_{\rm ghost} - S_{\rm ghost}^{\rm free}
  =  2{\cal J}\sum_{n,\mu} \tr \big\{\bar c(n)\, [D^-_\mu D^+_\mu -
    \Delta^-_\mu\Delta^+_\mu] \,c(n)\big\}\,.
\end{aligned}
\end{equation}
Moreover, extra vertices come from the terms 
\begin{equation}
\label{eq:int_action_AB}
\begin{aligned}
  S_{\rm A} &= {\cal J}a^2\sum_{n,\mu,\nu} \tr
  \big\{R_{\mu\nu}^{(1)}(n)f_{\mu\nu}(n)\big\} + 
  S_{\rm g.f.}^{(2)}
= {\cal J}a^2\sum_{n,\mu,\nu} \tr
  \big\{[R_{\mu\nu}^{(1)}(n)+\bar R_{\mu\nu}^{(1)}(n)]f_{\mu\nu}(n)\big\}\,,
 \\
  S_{\rm B} &= {\cal J}a^2\sum_{n,\mu,\nu} \tr
  \big\{R_{\mu\nu}^{(2)}(n)f_{\mu\nu}(n)\big\}\,, 
\qquad  S_{\rm T} =  -{\cal J}a^4\sum_{n,\mu,\nu}
  \frac{1}{(2\lambda_\mu\lambda_\nu)^2}\tr \big\{F_{\mu\nu}^{\,2}(n)
  f_{\mu\nu}^{\,2}(n)\big\}\,. 
\end{aligned}
\end{equation}
Explicitly, we have for $S_A$ and $S_B$ the
expressions\footnote{\label{foot:ho1} In
  these quantities one should in principle include 
  also the higher-order terms mentioned above in footnote
  \ref{foot:ho0}, by properly redefining $f_{\mu\nu}$.}
\begin{equation}
  \label{eq:SASB}
\begin{aligned}
  S_{\rm A} &= {\cal J}a^2\sum_{n,\mu,\nu}\frac{1}{2}\tr \big\{A_{\mu\nu}(n)
  f_{\mu\nu}(n)\big\}\,,\\
  A_{\mu\nu} &= 2i\left(2[\Q_\mu,\Q_\nu] +
  \frac{1}{\lambda_\mu}[\Q_\mu,D^+_\mu \Q_\nu]
  -\frac{1}{\lambda_\nu}[\Q_\nu,D^+_\nu \Q_\mu]
  \right. 
\left.
  -\frac{1}{2\lambda_\mu\lambda_\nu}[D^+_\mu \Q_\nu, D^+_\nu
    \Q_\mu]\right),\\
  S_{\rm B} &= {\cal J}a^2\sum_{n,\mu,\nu}\frac{1}{2}\tr \big\{B_{\mu\nu}(n)
  f_{\mu\nu}(n)\big\}\,,\\
  B_{\mu\nu} &=
  {i}\left(\frac{1}{\lambda_\mu}[\Q_\mu,D^+_\nu \Q_\mu] -
    \frac{1}{\lambda_\nu}[\Q_\nu, D^+_\mu \Q_\nu]\right).
\end{aligned}
\end{equation}
Notice that the terms $S_A$ and $S_{\rm T'}$ are odd in a given
component $\Q_\mu$ of the gluon field, while the other terms are
even. Since the propagator is diagonal, this implies~\cite{DG,Karsch,GAKA}
that\footnote{\label{foot:ho2} This clearly remains true also if 
  higher-order terms neglected in Eq.~\eqref{eq:bgplaq} are included in
  the definition of $f_{\mu\nu}$, see footnotes \ref{foot:ho0} and
  \ref{foot:ho1}. Since 
  $S_{\rm gluon}^{\rm int}$ is $\Oc(B)$, the only contribution of a
  higher-order term which should still be considered is that of the
  $\Oc(a^3 B\de B)$ term in Eq.~\eqref{eq:bgplaq} to $\la
  S_{\rm B} \ra_0 \propto \sum \tr \{\la B_{\mu\nu}\ra_0 t^a\}
  f^a_{\mu\nu}$; however, (global) background gauge invariance implies
  that $\tr\{\la B_{\mu\nu}\ra_0 t^a\}\propto B^a$, and so
  higher-order terms can be safely ignored.} $\la S_{\rm 
  A} \ra_0 = \la S_{\rm T'} \ra_0 = 0$, and also that  
$ \la S_{\rm gluon}^{\rm int} S_{\rm A} \ra_{0\,c}  =
\la S_{\rm A} S_{\rm B} \ra_{0\,c}=0$.

\subsubsection{The effective action}

Expanding now Eq.~\eqref{eq:G_BF1l3} up to terms quadratic in the
background f\mbox{}ield, we obtain the following expression for
$S_{\rm ef\mbox{}f}[B]$,  
\begin{equation}
\begin{aligned}
  \label{eq:effaction2}
  S_{\rm ef\mbox{}f}|_{\Oc(B^2),\Oc(g^0)} =~& 
  S_c + \frac{1}{2}\left(\la S_{\rm gluon}^{\rm int} \ra_0  -
    \frac{1}{2}\la \left(S_{\rm 
    gluon}^{\rm int}\right)^2 \ra_{0\,c} \right)  -
  \frac{1}{2}\la \left(S_{\rm A}\right)^2 \ra_{0\,c} \\ & -   \frac{1}{2}\la
  \left(S_{\rm B}\right)^2 
  \ra_{0\,c} + \la S_{\rm T} \ra_0
+ \left(\la S_{\rm B} \ra_0 - \la S_{\rm gluon}^{\rm int} S_{\rm
    B} \ra_{0\,c}\right) \\ \equiv~& S_c + \Delta S_{\rm g} + \Delta S_{\rm A} +
\Delta S_{\rm B} +
\Delta S_{\rm T} + \Delta S_{\rm gB}\,. \phantom{\frac{1}{2}}
\end{aligned}
\end{equation}
Here we have taken into account the remarks after Eq.~\eqref{eq:SASB},
and the fact that in four dimensions the ghost contribution exactly cancels
half of the gluon contribution from $S_{\rm
  gluon}^{\rm int}$~\cite{DG,HH,Karsch,GAKA}. Terms have been grouped so 
that each contribution is separately gauge-invariant~\cite{GAKA}.

\begin{table}
  \centering
  \begin{tabular}{c|l}
    term & $\Delta K_{\mu\nu}$ \\
    \hline\hline\\
   $\Delta S_{\rm g}$
    &  $-\displaystyle\frac{N_c}{3(4\pi)^2}\left[-\gamma + 
      \log \frac{1}{(aM)^2} + \frac{8}{3}\right] 
    - \frac{N_c}{3}[{\cal G}_{\mu\nu}(\lambda)-{\cal G}_{\mu}(\lambda)-{\cal
      G}_{\mu}(\lambda)+{\cal G}(\lambda)]$ \\ 
    \\   
    $\Delta S_{\rm A}$
    & $\displaystyle \frac{4N_c}{(4\pi)^2}\left[-\gamma + \log
      \frac{1}{(aM)^2} +2\right]
    +N_c\left[
      {\cal G}_{\mu\nu}(\lambda) - 2{\cal G}_\mu(\lambda) - 2{\cal
        G}_\nu(\lambda)  + 4{\cal G}(\lambda) 
    \right]$\\ \\
    $\Delta S_{\rm B}$
    &  $\displaystyle \frac{N_c}{4}
    \left[
      {\cal Z}(\lambda)\left(\frac{1}{\lambda_\mu^2} + \frac{1}{\lambda_\nu^2}
      \right) -
      \frac{{\cal Z}_\mu(\lambda)}{\lambda_\nu^2}
      -\frac{{\cal Z}_\nu(\lambda)}{\lambda_\mu^2}
    \right]$\\ 
    \\   
    $\Delta S_{\rm T}$
    &   $\displaystyle
    \frac{N_c^2-1}{2N_c}\left[
      \frac{{\cal Z}_\mu(\lambda)}{\lambda_\nu^2}
      +\frac{{\cal Z}_\nu(\lambda)}{\lambda_\mu^2}
    \right]$\\ 
    \\   
    $\Delta S_{\rm gB}$
    & 0 
  \end{tabular}
  \caption{Contribution of the various terms in Eq.~\eqref{eq:effaction2}
    to $K_{\mu\nu}$ in the effective action,
    Eq.~\eqref{eq:bfm_main}. 
  } 
\label{tab:coeff}
\end{table}

The evaluation of the various terms is performed generalising the
techniques developed in~\cite{GAKA} to the anisotropic case. Since such a
generalisation is straightforward, here we simply list the results,
giving in Tab.~\ref{tab:coeff} the contribution $\Delta{
  K}_{\mu\nu}$ of each term to the quantity ${K}_{\mu\nu}$ [see
Eqs.~\eqref{eq:bfm_main} and \eqref{eq:restore1}] in
front of $1/2\int d^4x\, \tr {\cal F}^{\,2}_{\mu\nu}(x)$. 
The relevant technical details can be found 
in the appendix of Ref.~\cite{GAKA}. Summing up, one obtains 
\begin{equation}
  K_{\mu\nu}(a,\lambda) = K^{div}(a) + {\cal K}_{\mu\nu}(\lambda) =  \beta_0\log
  \frac{1}{(aM)^2} + {\cal K}_{\mu\nu}(\lambda), 
\end{equation}
with $M$ a mass scale which sets the renormalisation point, 
$\beta_0$ the f\mbox{}irst coef\mbox{}f\mbox{}icient of the 
Yang-Mills $\beta$-function~\cite{tH,Gr,Po}, 
\begin{equation}
\label{eq:beta_func}
  \beta_0 = \frac{11}{3}\frac{N_c}{(4\pi)^2},
\end{equation}
and with ${\cal K}_{\mu\nu}$ f\mbox{}inite, $a$-independent
coef\mbox{}f\mbox{}icients, 
\begin{equation}
  \label{eq:K_app}
\begin{aligned}
  {\cal K}_{\mu\nu}(\lambda) =~& \frac{11}{3}\frac{N_c}{(4\pi)^2}\left[-\gamma +
    \frac{64}{33}\right] + N_c\left[
\frac{2}{3} {\cal G}_{\mu\nu}(\lambda) -\frac{5}{3}( {\cal
  G}_{\mu}(\lambda) + {\cal G}_{\nu})(\lambda) + 
   \frac{11}{3} {\cal G}(\lambda)\right] 
\\ 
&+ \frac{N_c}{4}\left[
{\cal Z}(\lambda)\left(\frac{1}{\lambda_\nu^2} + \frac{1}{\lambda_\mu^2}
\right) -
\frac{{\cal Z}_\mu(\lambda)}{\lambda_\nu^2}
-\frac{{\cal Z}_\nu(\lambda)}{\lambda_\mu^2}
\right]
  + \frac{N_c^2-1}{2N_c}\left[
\frac{{\cal Z}_\mu(\lambda)}{\lambda_\nu^2}
    +\frac{{\cal Z}_\nu(\lambda)}{\lambda_\mu^2}
\right]\,,
\end{aligned}
\end{equation}
where $\gamma\simeq 0.5772$ is the Euler--Mascheroni constant,
and ${\cal G}_{\mu\nu}$, ${\cal G}_{\mu}$,
${\cal G}$, ${\cal Z}_{\mu}$ and ${\cal Z}$  are functions of
$\{\lambda_\mu\}$ defined in terms of integrals involving the modified
Bessel functions of the first kind. Their precise form is not needed
for the analysis of the present Section, and can be found in 
Appendix \ref{app:GZanal}, Eqs.~\eqref{eq:integrals} and
\eqref{eq:integrals2}.  

To renormalise the theory and recover $O(4)$ invariance in the
continuum limit it is enough to set 
\begin{equation}
  \label{eq:renorm}
  \f{1}{g^2} = \f{1}{g^2_r(M)} + \beta_0\log
  \frac{1}{(aM)^2} = \beta_0\log\frac{1}{(a\Lambda)^2}
\,, \quad \f{\delta \beta_{\mu\nu}}{2N_c} = {\cal K}_{\mu\nu}\,.
\end{equation}
Here $\Lambda = M \exp\{-1/(2\beta_0 g_r^2(M))\}$ is a
renormalisation-group-invariant mass scale, whose value can be 
determined by comparing lattice results with experiments. 
Since a shift $\delta \beta_{\mu\nu}\to \delta
\beta_{\mu\nu} + \tilde\beta$ can always be reabsorbed in a
redefinition of $g$, any choice satisfying the set of conditions $\delta
\beta_{\mu\nu} - \delta \beta_{\rho\sigma} = {\cal K}_{\mu\nu} - {\cal
  K}_{\rho\sigma}$ will actually lead to restoration of $O(4)$
invariance at one-loop accuracy.\footnote{More
  generally, it is the ratios $\beta_{\mu\nu}/\beta_{\rho\sigma}$ that
will be constrained by the request of  restoration of $O(4)$
invariance, see the discussion in Section \ref{sec:anis_ren}.}
As we show in Appendix \ref{app:GZanal}, under a global rescaling
$\lambda_\mu\to\zeta\lambda_\mu$, ${\cal Z}$ and ${\cal Z}_\mu$ get a 
factor $\zeta^2$, ${\cal G}_{\mu\nu}$ and  ${\cal G}_{\mu}$ are
unchanged, and  ${\cal G}\to {\cal G} + \f{1}{(4\pi)^2}\log\zeta^2$, so
that overall ${\cal K}_{\mu\nu}\to {\cal K}_{\mu\nu} +
\beta_0\log\zeta^2$. Since the additive term can be cancelled by
$a\to\zeta a$, this means that the couplings $\beta_{\mu\nu}$ depend
on $a$ and $\lambda_\mu$ only through the combinations provided by the 
lattice spacings $a_\mu$, as they should. As we have already remarked,
to avoid redundancy one has to impose a condition on the
$\lambda_\mu$'s, like, e.g., setting $\lambda_\alpha=1$ for some $\alpha$, so
using one of the lattice spacings as reference length, or imposing the
symmetric condition $\prod_\alpha \lambda_\alpha =1$, thus using the
volume of the elementary cell to define $a$. 

We have compared our results with the ones available in the
literature for the isotropic~\cite{DG,HH,GAKA},
3+1~\cite{Karsch,GPvB,DHHS,Bur} and 2+2~\cite{Bur} anisotropic
cases.\footnote{In the 3+1 anisotropy class one
  lattice spacing differs from the other three, e.g.,
  $\lambda_4\neq\lambda_1=\lambda_2=\lambda_3$,  while in the 2+2
  class the lattice spacings are equal pairwise, e.g.,
  $\lambda_4=\lambda_1\neq\lambda_2=\lambda_3$.} In  
particular, we have successfully checked that in the isotropic case we
recover the result of~\cite{GAKA}, and we have compared the
differences of $\delta\beta_{\mu\nu}$ with the ones reported
in Ref.~\cite{Bur} for the 3+1 and 2+2 cases. While there is full agreement
for the 3+1 case, we found a discrepancy in the analytic expression of
one of the two independent differences in the 2+2 case.\footnote{
In the notation of Ref.~\cite{Bur}, the discrepancy is in
$\eta^{(1)}_{ff}-\eta^{(1)}_{cf}$, in particular in the coefficients
of the quantities ${\cal B}_\xi^c(2,1)$ and ${\cal B}_\xi^f(2,1,1)$,
for which we find respectively $\f{N_c}{2}(\f{1}{\xi^2}+\f{5}{3\xi^4})$
and $\f{N_c}{6}(\f{1}{2}+\f{1}{\xi^2})$.} On the other hand, the
numerical values also reported in Ref.~\cite{Bur} agree 
with ours. It has to be noted that the analytic result reported
in Ref.~\cite{Bur} for that difference does not vanish when there 
is no anisotropy, as it should, so most likely it contains some
misprint.  

For future utility, we report the lowest-order approximation for the
expectation value $\la{\cal P}_{\mu\nu}\ra$ of the plaquette
terms. Setting 
$U_\mu(n)=e^{i\f{g}{\lambda_\mu}\Q_\mu}$ and expanding in $g$, one finds
\begin{equation}
  \label{eq:plaq_1}
  \la{\cal P}_{\mu\nu}\ra =
  \f{g^2}{2N_c}\f{1}{\lambda_\mu^2\lambda_\nu^2}\la\tr
  F_{\mu\nu}^2\ra_0 + \Oc(g^3) \,, 
\end{equation}
where $F_{\mu\nu}=\Delta^+_\mu \Q_\nu - \Delta^+_\nu \Q_\mu$ [see
Eq.~\eqref{eq:FRdef}], and $\la\dots\ra_0$ has been defined in
Eq.~\eqref{eq:ave}. A straightforward calculation yields 
\begin{equation}
  \label{eq:plaq_2}
 \la   {\cal P}_{\mu\nu}\ra  
= g^2\f{N_c^2-1}{2N_c}\left[\f{{\cal Z}_\mu(\lambda)}{\lambda_\nu^2} +
      \f{{\cal Z}_\nu(\lambda)}{\lambda_\mu^2}\right]
    + \Oc(g^3) \,.
\end{equation}

\newpage
\section{Analytic continuation in the nonperturbative approach to
  soft high-energy scattering}
\label{sec:AC}

In this Section we use the results of Section \ref{sec:anis_ren} in the
context of the nonperturbative approach to soft high-energy
scattering. After a brief review of this approach (the interested
reader can confer
Refs.~\cite{Nachtmann91,DFK,Nachtmann97,BN,Dosch,LLCM1,pomeron-book,reggeon} 
for a more detailed discussion), we discuss its formulation on a
Euclidean anisotropic lattice, and we refine the arguments of
Ref.~\cite{GM2009} on the analytic continuation back to Minkowski
spacetime.

\subsection{Euclidean approach to soft high-energy scattering}
\label{sec:AC_app}

Soft high-energy scattering is characterised by small transferred momentum
squared $t$, $|t|\lesssim 1~{\rm GeV}^2$, and very large total
center-of-mass energy squared $s$, $s\gg 1~{\rm GeV}^2$. In the
approach of Ref.~\cite{Nachtmann91}, hadronic scattering amplitudes in
the soft high-energy regime can be obtained from partonic
scattering amplitudes after folding with appropriate hadronic wave
functions. In particular, for meson-meson
scattering the basic quantity is the scattering amplitude of two
colourless transverse dipoles, which in the soft high-energy regime is
given in impact-parameter space by the correlation function of two
rectangular Minkowskian Wilson loops~\cite{DFK,Nachtmann97}. These
Wilson loops are computed on the paths described by the 
classical trajectories of the dipoles, so forming a large hyperbolic
angle $\chi$ in the longitudinal plane, and are cut at proper times
$\pm T$ for infrared regularisation purposes~\cite{Verlinde}. In turn,
their (Minkowskian) correlation function is obtained after analytic
continuation in the angular variable and in the length of the loops
from the correlation function of two Euclidean Wilson loops of length
$2T$ forming an angle $\theta$ in the longitudinal Euclidean
plane~\cite{GM2009,analytic1,analytic2,analytic3,Meggiolaro05,crossing1,crossing2}.  
This approach can be generalised to describe scattering processes
involving
baryons~\cite{Nachtmann91,DFK,Nachtmann97,BN,Dosch,LLCM1,pomeron-book,RD}. As
the constructions and the arguments of this Section are easily
adapted to this case, we restrict the discussion to meson-meson
(dipole-dipole) scattering for simplicity. 

\begin{figure}[t]
  \centering
  \includegraphics[width=0.4\textwidth]{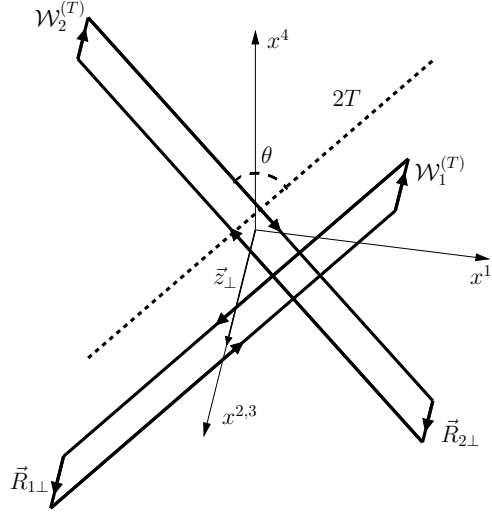}
  \caption{The Euclidean Wilson loops $\W_{1}^{(T)}$ and $\W_{2}^{(T)}$, defined in
    Eq.~\eqref{eq:trajE}.} 
  \label{fig:0}
\end{figure}
The relevant Euclidean correlator is given
by\footnote{Here and in the following we denote by $\vec v_\perp$ a
  two-dimensional vector in the transverse plane.}
\begin{equation}
  \label{eq:E_corr}
  {\cal G}_E(\theta,T;\vec
  z_\perp;\vec{R}_{1\perp},f_1;\vec{R}_{2\perp},f_2) = \f{\la \W_1^{(T)}
    \W_2^{(T)}\ra_E}{\la \W_1^{(T)} \ra_E \la 
    \W_2^{(T)}\ra_E} -1\,,
\end{equation}
where $\la\ldots\ra_E$ denotes the expectation value in the sense of
the Euclidean functional integral, $\vec z_\perp$ is the
impact-parameter distance between the dipoles, and $\vec{R}_{i\perp}$
and $f_i$ are the transverse size of the dipoles and the longitudinal
momentum fraction of the quarks in the two mesons, respectively
(``dipole variables''). 
The Wilson loops $\W_{1,2}^{(T)}$ are computed on the following paths
(see Fig.~\ref{fig:0}),
\begin{equation}
  \begin{aligned}
    {\cal C}^{\,(T)}_1 &: {X}_{E1}^{\pm}(\tau) =  \pm
    u_1 \tau + z
    + f^{\pm}_1 R_{1}
    =\pm u_1 \tau + d_1^\pm
\, , \\
{\cal C}^{\,(T)}_2 &: {X}_{E2}^{\pm}(\tau) =
\pm u_2 \tau 
+ f^{\pm}_2 R_{2}
 =\pm u_2 \tau + d_2^\pm\,,
  \end{aligned}
\label{eq:trajE}
\end{equation}
with $\tau\in [-T,T]$, and closed by straight-line paths in the
transverse plane at $\tau=\pm T$. The four-vectors $u_{1,2}$ are
chosen to be $u_{1,2}=(\pm\sin\frac{\theta}{2}, \vec{0}_{\perp},
\cos\frac{\theta}{2})$, $\theta$ being the angle formed by the two
trajectories, i.e., $u_1\cdot u_2 = \cos\theta$. Moreover,
$R_{i} = (0,\vec{R}_{i\perp},0)$, $z = (0,\vec{z}_{\perp},0)$ and
$f^+_i \equiv 1-f_i$, $f^{-}_i \equiv -f_i$, with $f_i\in[0,1]$.
The Minkowskian correlation function is obtained from
Eq.~\eqref{eq:E_corr} by means of analytic continuation as
follows~\cite{GM2009,Meggiolaro05}, 
\begin{equation}
  \label{eq:an_cont_corr}
  {\cal G}_M(\chi,T;\vec
  z_\perp;\vec{R}_{1\perp},f_1;\vec{R}_{2\perp},f_2) =
  {\cal G}_E(-i\chi,iT;\vec
  z_\perp;\vec{R}_{1\perp},f_1;\vec{R}_{2\perp},f_2)\,.
\end{equation}
Physical amplitudes are finally obtained from ${\cal G}_M$ in the
limit $T\to\infty$, and for asymptotically large $\chi\sim\log s$. 
It is worth mentioning that combining Eq.~\eqref{eq:an_cont_corr} with
the $O(4)$ symmetry of the Euclidean theory one obtains the following 
crossing-symmetry relations~\cite{crossing1,crossing2},
\begin{equation}
  \label{eq:cross_rel}
  {\cal G}_M(\chi,T;\vec
  z_\perp;\vec{R}_{1\perp},f_1;-\vec{R}_{2\perp},1-f_2) =
  {\cal G}_M(i\pi-\chi,T;\vec
  z_\perp;\vec{R}_{1\perp},f_1;\vec{R}_{2\perp},f_2)\,,
\end{equation}
which allow us to relate the scattering amplitudes in the direct
(meson-meson) and crossed (meson-antimeson) channels.

The analytic continuation relation, Eq.~\eqref{eq:an_cont_corr}, has
allowed studies of the correlators through nonperturbative Euclidean
techniques~\cite{sigtot,LLCM2,ILM,JP,JP2,Janik,GP2010,GM2008,GM2010,
GMM,sigtot_comments}. For a brief review of the older results and a
comparison to lattice data cf.~\cite{GM2008,GM2010,GMM}.

\subsection{Anisotropic lattice formalism}
\label{sec:AC_lat}

It is well known that the functional integral needs to be regularised
to become a well-defined mathematical object. Furthermore, the
analytic continuation relation Eq.~\eqref{eq:an_cont_corr} is
meaningful only if a sufficiently wide analyticity domain exists. The
first issue can be dealt with by discretising the theory on a lattice,
so that the relevant Wilson loop correlator can then be computed
nonperturbatively, for example by means of numerical simulations,  
using off-axis operators to approximate the continuum Wilson
loops. Numerical simulations using an isotropic lattice have been
reported in Refs.~\cite{GM2008,GM2010,GMM}. Unfortunately, only a
discrete set of angles is accessible in this case; furthermore, for
each angle one has to use a different off-axis Wilson loop, which
makes the angular dependence even less analytically
controllable. Since our purpose here is to study the analytic
dependence on $\theta$ and $T$, it is more convenient to use an
appropriate anisotropic lattice keeping fixed the Wilson-loop
operator, which allows us to expose the dependence on the relevant
variables in the action. In this way we make the functional integral a 
well-defined object, and at the same time we can study the analyticity
domain of the correlator.

To avoid complications related to the well-known difficulties in
treating fermions on the lattice, in this study we consider the {\it
  quenched} approximation of QCD, i.e., the pure-gauge theory case. We
hope to return in a future paper on the inclusion of fermionic
effects, which may be more important than usually expected for soft
high-energy processes (see Refs.~\cite{sigtot,sigtot_comments}).

A good choice is to use the anisotropic action discussed previously,
Eq.~\eqref{eq:action}, taking the anisotropy parameters to be such that
the long sides of the Wilson loops lie in a lattice plane at
$45^\circ$ from two of the lattice axes, and are of fixed length. This
amounts to set 
\begin{equation}
  \label{eq:anis_param}
  \lambda_4(\theta,\barT) = \frac{1}{\sqrt{2}\barT
    \cos\frac{\theta}{2} } \,, 
  \quad
  \lambda_1(\theta,\barT) = \frac{1}{\sqrt{2}\barT \sin\frac{\theta}{2}} \,,
  \quad 
  \lambda_2(\theta,\barT) = \lambda_3(\theta,\barT) = 1,
\end{equation}
where $\barT\equiv T/T_0$ with $T_0$ some fixed length, and $\theta$ is
restricted to $\theta\in (0,\pi)$ without loss of 
generality~\cite{crossing1}. This yields for the plaquette coefficients
\begin{equation}
  \label{eq:tree_lev_coeff}
  \begin{aligned}
    {C}_{41}(\theta,\barT) & = 
    \f{1}{{C}_{23}(\theta,\barT)}  =
    \dfrac{1}{\barT^2\sin\theta}\,, \\
    {C}_{42}(\theta,\barT) & = {C}_{43}(\theta,\barT) =
    \f{1}{{C}_{12}(\theta,\barT)} 
    = \f{1}{{C}_{13}(\theta,\barT)}
    =  \tan\f{\theta}{2}\,.
  \end{aligned}
\end{equation}
Notice that the following relations hold,
\begin{equation}
  \label{eq:relations}
      \lambda_4^2(\theta,\barT)=
      \frac{C_{42}(\theta,\barT)}{C_{23}(\theta,\barT)}\,,  
\quad
  \lambda_1^2(\theta,\barT)=  \frac{C_{12}(\theta,\barT)}{C_{23}(\theta,\barT)}\,,
\quad
{\cal J}(\theta,\barT)=\barT^2\sin\theta ={C}_{23}(\theta,\barT) \,. 
\end{equation}
The action defined by Eq.~\eqref{eq:action}, with anisotropy parameters
Eq.~\eqref{eq:anis_param}, will be denoted by
$S[U;\theta,\barT]$, and the corresponding expectation value will be
denoted by $\la\ldots\ra_{\theta,\bar T}$.

\begin{figure}[t]
  \centering
  \includegraphics[width=0.5\textwidth]{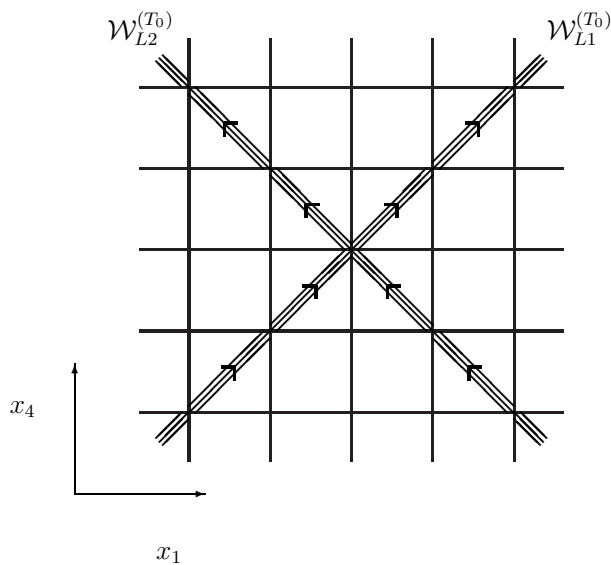}
  \caption{The ``tilted'' lattice Wilson loops $\W_{L1}^{(T_0)}$ and
    $\W_{L2}^{(T_0)}$, 
    Eqs.~\eqref{eq:lat_WL} and \eqref{eq:lat_WL2}, projected on the
    longitudinal plane.} 
  \label{fig:1}
\end{figure}

The lattice Wilson loops are defined as
\begin{equation}
  \label{eq:lat_WL}
  \W_{Li}^{(T_0)} = \f{1}{N_c}\tr \{W_i^+ H_i^+ W_i^-{}^\dag H_i^-{}^\dag\}\,,
\end{equation}
where the ``tilted'' Wilson lines $W_i^\pm$ are defined as (see
Fig.~\ref{fig:1})  
\begin{equation}
  \label{eq:lat_WL2}
  W_i^\pm = \prod_{j=-t_0}^{t_0-1} {\cal U}^{(i)}(jv_{1,2} + d_{Li}^\pm)\,,
\end{equation}
where $v_{1,2}=(\pm 1,0,0,1)$, $t_0=\f{T_0}{a\sqrt{2}}$ 
with $t_0\in\mathbb{N}$, and $d_{Li}^\pm=d_{i}^\pm/a$ denotes the
transverse position in lattice units, see Eq.~\eqref{eq:trajE}, while  
$H_i^\pm$ are the appropriate Wilson lines made of the usual
link variables in the transverse plane, closing the loops.\footnote{One can 
properly choose $\lambda_{2,3}$ and use ``tilted'' links also in the
transverse plane. This would however leave the discussion and the
conclusions of this Section unchanged.} 
It is clear that $\sqrt{2}t_0=\f{T_0}{a}$ is the distance in lattice
units between the center of a long side of the loop and its endpoints,
i.e., loosely speaking, the half-length in lattice units of the Wilson
loops.  The ``tilted links'' 
${\cal U}^{(i)}(n)$ are appropriate 
functionals ${\cal U}^{(i)}[U;n]$ of the lattice links, 
which in the continuum limit have to satisfy\footnote{The factor
  in front of the square brackets takes into account that the diagonal of
  a plaquette in the longitudinal plane has length 
  $\sqrt{a_4^2+a_1^2}= \sqrt{2}a\barT$. 
  Notice that we are using path-ordered Wilson loops, as it is customary
  on the lattice, rather than the time-ordered Wilson loops appearing in
  the formulae for the scattering amplitudes (see,
  e.g., Refs.~\cite{reggeon,GM2008}). This has no consequence on the
  results, as the theory is invariant under charge conjugation, and so
  under reversing the loop orientation.
} [see Eq.~\eqref{eq:links_a}]
\begin{equation}
  \label{eq:tilted_link}
  \begin{aligned}
    {\cal U}^{(1)}(n) &=
\I +    
ia{\barT\sqrt{2}}\left[\cos{\textstyle\f{\theta}{2}}
A_4(x(n)) 
 + 
    \sin{\textstyle\f{\theta}{2} }
A_1(x(n))\right]
    +\Oc(a^2)
\,, \\
    {\cal U}^{(2)}(n) &=
\I +    
ia{\barT\sqrt{2}}\left[\cos{\textstyle\f{\theta}{2}}
A_4(x(n)) 
- 
    \sin{\textstyle\f{\theta}{2} }
A_1(x(n))\right]
    +\Oc(a^2)
\,,
    \end{aligned}
\end{equation}
and which under a gauge transformation behave
as
\begin{equation}
  \label{eq:tilt_gtransf}
  \begin{aligned}
 {\cal U}^{(1)}(n)\to G(n){\cal U}^{(1)}(n)G^\dag(n+\hat 4 +\hat
1) \,,   \\
 {\cal U}^{(2)}(n)\to G(n){\cal U}^{(2)}(n)G^\dag(n +\hat 4 -\hat 1) \,. 
  \end{aligned}
\end{equation}
\begin{figure}[t]
  \centering
  \includegraphics[width=0.6\textwidth]{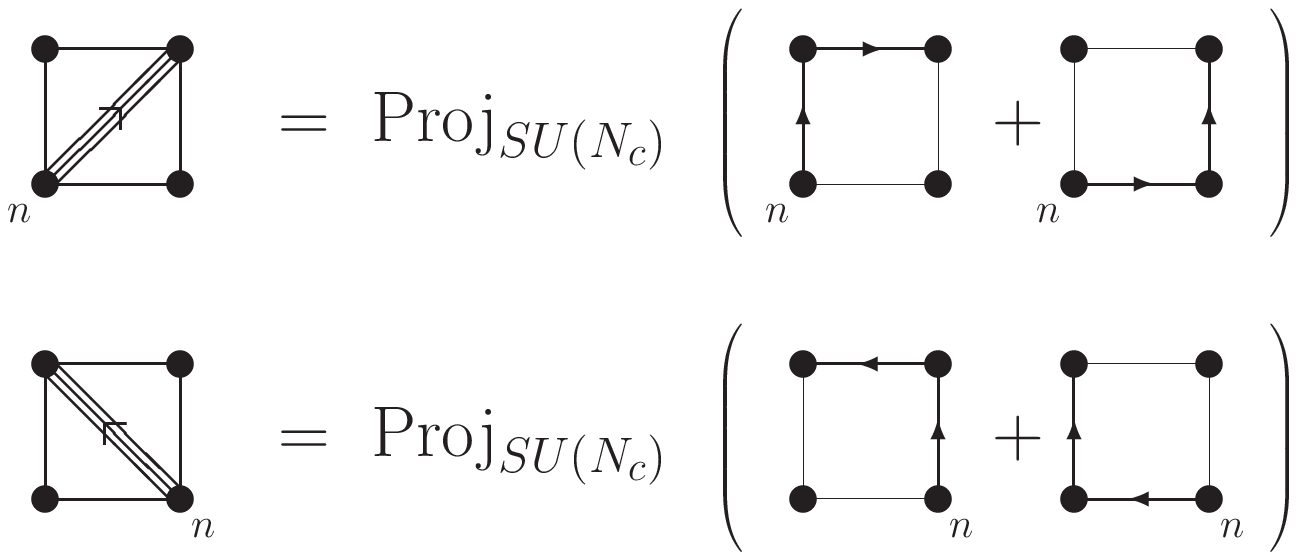}
  \caption{The ``tilted'' links of Eq.~\eqref{eq:tilt_link_unit}, 
    built from the shortest paths 
    connecting opposite corners of a plaquette.}
  \label{fig:2}
\end{figure}
The simplest possibility in
building ${\cal U}^{(1,2)}$ is obviously to use a combination of the gauge
transporters along the two shortest paths connecting opposite corners
of an elementary plaquette, namely
\begin{equation}
  \label{eq:gauge_transp}
  \begin{aligned}
    U^{(1)}_1(n) &= U_4(n)U_1(n+\hat 4)\,, &&&     
    U^{(1)}_2(n) &= U_1(n)U_4(n+\hat 1)\,, \\
    U^{(2)}_1(n) &= U_4(n)U_1^\dag(n+\hat 4-\hat 1)\,, &&&
    U^{(2)}_2(n) &= U_1^\dag(n-\hat 1)U_4(n-\hat 1) \,.
  \end{aligned}
\end{equation}
It is convenient to adopt a definition of ${\cal U}^{(j)}$ which is
symmetric under the exchange $U^{(j)}_1(n)\leftrightarrow
U^{(j)}_2(n)$. A viable choice is (see Fig.~\ref{fig:2}) 
\begin{equation}
  \label{eq:tilt_link_unit}
{\cal U}^{(j)}(n) = {\rm Proj}_{SU(N_c)}\left[ U^{(j)}_1(n) +
  U^{(j)}_2(n)\right]\,, 
\end{equation}
with ${\rm Proj}_{SU(N_c)}$ denoting the projection on $SU(N_c)$. 
This symmetry requirement comes out naturally if we want that the
Wilson loop correlator satisfies on the lattice the same ``crossing 
property''~\cite{crossing1,crossing2} that it satisfies in the
continuum. It is easy to show that in the continuum the correlation
function of the two Wilson loops $\W_{1,2}^{(T)}$, defined 
in Eq.~\eqref{eq:trajE}, at angle $\pi-\theta$ is equal to the
correlation function of $\W_{1,2}^{(T)}$ at angle $\theta$ but with
the orientation of one of the loops being reversed. In formulae,
\begin{equation}
  \label{eq:simple_crossing}
 \la \W_1^{(T)} \W_2^{(T)}\ra_E|_{\theta=\pi-\vartheta}=\la \W_1^{(T)}
 \W_2^{(T)\,*}\ra_E|_{\theta=\vartheta}=\la \W_1^{(T)\,*}
 \W_2^{(T)}\ra_E|_{\theta=\vartheta} \,. 
\end{equation}
In order to impose this symmetry on the lattice, let us first notice
that the anisotropic lattice action defined by
Eqs.~\eqref{eq:action} and \eqref{eq:anis_param} is invariant under 
the transformation $U=\mathbf{\Xi}  U^\Xi$  
acting on the links, defined by 
\begin{equation}
  \label{eq:trans_cross}
  \begin{aligned}
    &    U_4(n)=U^\Xi_1(n^\Xi)\,,\quad U_1(n)=U^\Xi_4(n^\Xi)\,,\quad
    U_{2,3}(n)=U^\Xi_{2,3}(n^\Xi)\,,\\
    &    n^\Xi_4 = n_1\,,\quad n^\Xi_1 = n_4\,,\quad n^\Xi_{2,3} = n_{2,3}\,,
  \end{aligned}
\end{equation}
if at the same time one also sends $\theta\to \pi - \theta$. Indeed, it
suffices to verify that $C_{42}(\pi-\theta,\barT)=C_{12}(\theta,\barT)$ and
$C_{41}(\pi-\theta,\barT)=C_{41}(\theta,\barT)$ [see
Eq.~\eqref{eq:tree_lev_coeff}]. Consequently, the one-loop corrections
${\cal K}_{\mu\nu}$ will transform in the same way as $C_{\mu\nu}$, as
can be also verified explicitly. 
We have then $S[\mathbf{\Xi} U;\theta,\barT] =
S[U;\pi-\theta,\barT]$, and since the integration measure is clearly 
invariant, the expectation value of some observable
$\Oc[U]$ satisfies $\la\Oc[U]\ra_{\pi-\theta,\bar T}=\la\Oc[\mathbf{\Xi}
U]\ra_{\theta,\bar T}$. In order to maintain the ``crossing property''
also on the lattice, the ``tilted links'' must therefore transform as
\begin{equation}
  \label{eq:tlink_crtrasf}
 {\cal U}^{(1)}[\mathbf{\Xi}U;n^\Xi]= {\cal U}^{(1)}[U;n]\,, \quad 
{\cal U}^{(2)}[\mathbf{\Xi}U;n^\Xi]= {\cal U}^{(2)}{}^\dag[U;n-\hat 4 +\hat 1] \,.
\end{equation}
One can then readily show that the definition
Eq.~\eqref{eq:tilt_link_unit} satisfies the properties
Eq.~\eqref{eq:tilted_link}, Eq.~\eqref{eq:tilt_gtransf} and 
Eq.~\eqref{eq:tlink_crtrasf}.
In Appendix \ref{app:abelian} we show that using
Eq.~\eqref{eq:tilt_link_unit} in the case of the compact $U(1)$ gauge
theory one correctly recovers the continuum result of Ref.~\cite{Meggiolaro05}
in the weak-coupling limit.

One can then define the relevant Euclidean correlator as the continuum
limit of the appropriate lattice correlator,
\begin{equation}
  \label{eq:WLC_def}
  \begin{aligned}
&    {\cal G}_E(\theta,T=T_0\barT) = \lim_{a\to 0,V\to\infty} {\cal
  G}_L(\theta,T_0,\barT;a,V)\,, \\
& {\cal G}_L(\theta,T_0,\barT;a,V) \equiv
 \f{\la \W_{L1}^{(T_0)}
      \W_{L2}^{(T_0)}\ra_{\theta,\bar T}}{\la \W_{L1}^{(T_0)} \ra_{\theta,\bar T} \la
      \W_{L2}^{(T_0)}\ra_{\theta,\bar T}} -1  \,,
  \end{aligned}
\end{equation}
where $V$ is the lattice volume, and we have dropped the dependence on
the impact parameter and on the dipole variables, since they play
no role in the following.  

\subsection{Analytic continuation}
\label{sec:AC_ac}

We now argue that ${\cal G}_E(w,T)$ is analytic in a complex domain ${\cal
  D}$ which makes the analytic continuation relations
Eq.~\eqref{eq:an_cont_corr} meaningful. Here $w$ and $T$ are now {\it
  complex} variables, which we parameterise as $w=\theta-i\chi$, with
real $\theta,\chi$, and $T=T_0 \bar T = T_0 |\bar T|
e^{i\f{\varphi}{2}}$, with 
$\varphi\in(-2\pi,
2\pi]$. Since one has to take two possibly dangerous limits, i.e., the
infinite-volume limit and the continuum limit, which currently are not
under full theoretical control, our argument is not
rigourous. Nevertheless, a few reasonable technical assumptions are
sufficient to complete the proof. 

The first thing to check is that the couplings, $\beta_{\mu\nu}(w,\barT)$, and
the plaquette coefficients, $C_{\mu\nu}(w,\barT)$, are analytic functions of
$w$ and $\barT=|\bar T| e^{i\f{\varphi}{2}}$. This is obvious at tree
level, since $\beta_{\mu\nu}=2N_c/g^2$ and the only singular points of
$C_{\mu\nu}$ are $w=n\pi$ with $n\in\mathbb{Z}$, and $\bar T
=0$. Analyticity of the one-loop corrections ${\cal
  K}_{\mu\nu}(w,\barT)$, and so of $\beta_{\mu\nu}(w,\barT)$ at the 
one-loop level, is studied in Appendix \ref{app:GZanal}. 

The next step is to require that the theory has the desired continuum
limit. This requires positivity of the real part of the action to
guarantee convergence. The tree-level convergence conditions have been
discussed in Ref.~\cite{GM2009}, and read
\begin{equation}
  \label{eq:conv_cond}
  \Re {C}_{\mu\nu}(w,\barT) > 0 \quad \forall \mu,\nu\,.
\end{equation}
These conditions define a complex domain ${\cal  D}$ which has been
fully worked out in Ref.~\cite{GM2009}. Although its detailed form
will not be used here, it is worth mentioning that ${\cal D}$ is
defined only in terms of the complex angle $w$ and of $\varphi$, 
i.e., $|\barT|$ is not restricted (except for asking $|\bar T|\neq
0$). 
The Euclidean region corresponds to $\theta\in (0,\pi)$,
$\chi=0$, $\varphi =0$. The Minkowskian region $\theta=0$, $\chi>0$,
$\varphi =\pi$ lies at the boundary of ${\cal D}$, and so does also
the ``crossed'' Minkowskian region $\theta=\pi$, $\chi<0$, $\varphi
=\pi$; we will refer to these as the ``physical'' boundaries of ${\cal
  D}$. Notice that both in the Euclidean and in the Minkowskian
regions the restrictions on the angular variables do not lead to any
loss of information~\cite{crossing1}. 
As it is shown in details in Appendix \ref{app:GZanal}, the one-loop
corrections ${\cal K}_{\mu\nu}(w,\barT)$ are analytic in ${\cal
  D}$. For small enough lattice spacing, the one-loop corrections will
therefore not spoil the positivity of the real part of the action
enforced at tree level, for any choice of parameters in 
a compact subdomain of ${\cal  D}$.

At finite volume and finite lattice spacing, and at one-loop accuracy,
we have therefore proved that the relevant correlators are analytic
functions in a domain ${\cal  D}$, within which  positivity of the
real part of the action is guaranteed. This domain of analyticity will
survive the infinite volume limit if the convergence is
uniform. Proving this is currently out of reach. However, if a lattice
system has short-range interactions, then correlation functions of
operators localised in some finite region ${\cal R}$ of spacetime will
become insensitive to the lattice size when this exceeds the size of
${\cal R}$ by a few correlation lengths. Notice that $T_0$ is fixed, so
that our operators are indeed localised. If interactions remain
short-ranged throughout ${\cal  D}$, then it is enough to take the
lattice size required by the largest correlation length 
(within some compact subdomain of ${\cal  D}$) to make finite-size
corrections uniformly negligible. This essentially amounts to assuming 
that the theory remains confining as one moves in ${\cal D}$. Although
we cannot prove this, we find it plausible: for example, it is
easy to see that it is true at strong coupling by means of a character
expansion.   

At this point one has to take the continuum limit. This limit is
expected to exist and be finite within ${\cal  D}$ (again, a rigorous
proof is out of question). In particular, Wilson-loop operators
renormalise multiplicatively~\cite{DV,BNS}, so that the normalised
correlation function appearing in Eq.~\eqref{eq:WLC_def} does not
require any further renormalisation on top of the renormalisation of
the couplings in the action, discussed in the previous Section. A
rigorous proof of uniform convergence is currently out of reach;
however, deviations from the continuum limit are expected to be of
order $\Oc(a)$, independently of $w$ and $\barT$, 
and in this case it is possible to make them uniformly negligible.  

The conclusion, within the present accuracy, is that $G_E$ is analytic
in the complex domain ${\cal D}$, which, as shown in Ref.~\cite{GM2009}, is
sufficiently wide to make the analytic continuation relation
Eq.~\eqref{eq:an_cont_corr} and the crossing-symmetry relations
Eq.~\eqref{eq:cross_rel} fully meaningful. 

As it was implicit in the discussion above, singularities in the
correlator may develop at the boundaries of ${\cal D}$. As the
analytic continuation Eq.~\eqref{eq:an_cont_corr} is formally
equivalent to the usual definition of the Minkowskian correlator
making use of the ``$-i\varepsilon$'' prescription~\cite{crossing1}, no
singularities are expected at the ``physical'' boundaries of the
domain. The anisotropic action itself is singular at 
$\theta=0,\pi$, $\chi=0$  and $\theta=0,\pi$, $\chi=\infty$, 
but as this is an artifact of the construction it
is not clear if true singularities are present there. At finite $T$
(i.e., for Wilson loops of finite physical length), no
singularity is expected in the Euclidean correlator ($\chi=0$) also at
$\theta=0,\pi$; however, 
as $T\to\infty$, a true singularity is expected to appear there, which
has its physical origin in the relation between the correlator
Eq.~\eqref{eq:E_corr} at $\theta=0,\pi$ and the static dipole-dipole
potential~\cite{Pot1,Pot2,Pot3,Pot4}. This is also supported by
numerical results~\cite{GM2008,GM2010}. On the other hand, the points
$\theta=0$, $\chi=\infty$ and $\theta=\pi$, $\chi=-\infty$, in the
limit $T\to\infty$, are 
the ones actually relevant to soft scattering at asymptotically high
energy, where the approach initiated by Ref.~\cite{Nachtmann91}
applies. A better understanding of the correlator near these points 
would help in the study of the asymptotic high-energy behaviour of
scattering amplitudes and total cross sections. 
In particular, in order to establish that the expressions for the
scattering amplitudes derived in this approach satisfy unitarity,  it
is crucial to show that for vanishing $\theta$ and large $\chi$ 
the correlator is a properly bounded function of the impact parameter
and of the dipole variables. 
Furthermore, the existence (or not) of the strict $\chi\to\infty$
limit at fixed impact parameter, and the properties of the correlator
in this limit, are closely connected to the issue of universality
of hadronic total cross sections observed in experiments (see, e.g.,
Refs.~\cite{uni-st1,uni-st2} and references therein). For more details
on these problems, we invite the interested reader to confer
Ref.~\cite{sigtot}.  

\begin{figure}[t]
  \centering
  \includegraphics[width=0.6\textwidth]{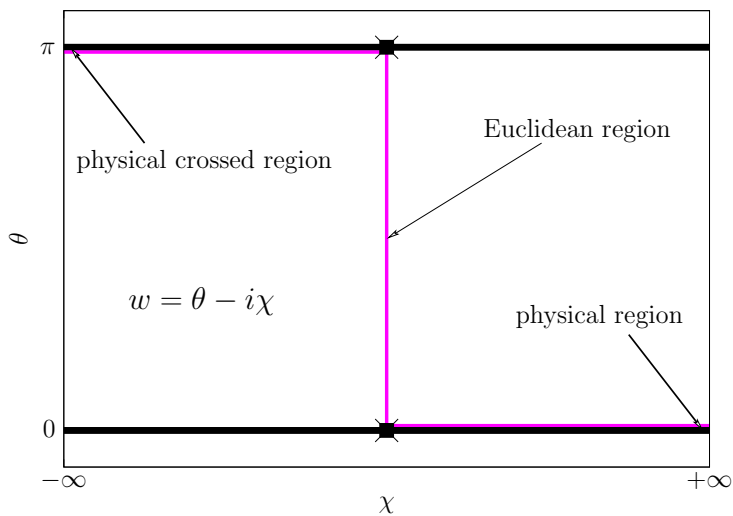}
  \caption{Analyticity domain of the Wilson-loop correlator with the
    infrared cutoff removed, ${\cal C}_E$ [see
    Eq.~\eqref{eq:an_cont_corr_C}]. The solid black lines indicate the
    boundaries of the domain, and crosses signal the singularities.} 
  \label{fig:3}
\end{figure}

Other singularities could appear when $|T|\to 0$ or
$|T|\to\infty$. Working at fixed $T_0$, this corresponds to
$|\barT|\to 0$ or $|\barT|\to\infty$, which are again singular points 
of the anisotropic action. However, since the analytic continuation
to the ``physical'' boundaries requires only the phase of $\barT$ to
be changed, one can take as well $T_0=|T|$ and
$\barT=\exp\{i\f{\varphi}{2}\}$, 
and study the two limits above by changing the length of the
``tilted'' Wilson loops.\footnote{In the continuum limit the choice of
  $T_0$ should be irrelevant, as long as it is compensated by the
  appropriate choice of $\barT$.} The above limits therefore
correspond to the limit of ``tilted'' Wilson loops of vanishing or
infinite length. In the first case no 
singularity is expected; in any case this limit is irrelevant for our
purposes. On the other hand, the limit of infinite length is the one
entering the physical scattering amplitudes. In this case, the
short-ranged nature of strong interactions (which is assumed to remain
unchanged throughout the analyticity domain) implies that distant
parts of the two Wilson loops do not ``feel'' each other, i.e.,
those parts of the loops that lie beyond a certain distance from the
centers interact mutually only very weakly, and essentially contribute
only to the self-interaction of the loops. These contributions are
cancelled by the normalisation factors, so that the correlator becomes
basically insensitive to the length of the loops beyond some critical
value, and a finite limit $|T|\to\infty$ is therefore expected. In the
Euclidean case, this has already been checked on the lattice,
although in an isotropic setting~\cite{GM2008}. As discussed
in Ref.~\cite{GM2009}, the boundedness and the analyticity properties
of the correlator as a function of $T$ imply through the
Phragm\'en-Lindel\"of theorem (see, e.g., Ref.~\cite{Tit}) that the
analytic continuation to Minkowski spacetime and the infinite-length
limit commute. Setting ${\cal C}_{E,M}=\lim_{T\to\infty} {\cal
  G}_{E,M}$, this means that one can obtain the physical correlator by
means of an analytic continuation in the angular variable only, i.e.,
\begin{equation}
  \label{eq:an_cont_corr_C}
  {\cal C}_M(\chi;\vec
  z_\perp;\vec{R}_{1\perp},f_1;\vec{R}_{2\perp},f_2) =
  {\cal C}_E(-i\chi;\vec
  z_\perp;\vec{R}_{1\perp},f_1;\vec{R}_{2\perp},f_2)\,.
\end{equation}
The analyticity domain for ${\cal C}_{E}(w=\theta-i\chi)$, already
discussed in Ref.~\cite{GM2009}, is clearly not changed by one-loop
corrections, and it is simply the strip $\theta\in (0,\pi)$,
$\chi\in\mathbb{R}$, shown in Fig.~\ref{fig:3}. 

\secspace
\section{Longitudinally rescaled action}
\label{sec:2dmodel}

The results of Section~\ref{sec:anis_ren} can be used to obtain some
insight in the 
approach to high-energy scattering based on longitudinally rescaled
actions~\cite{Verlinde,Arefeva1,Arefeva2,Orland:2008vg,Orland,Cubero:2011ut,
Cubero:2012nw}. The physical idea behind this approach is that
in high-energy scattering processes the longitudinal directions appear
highly Lorentz-contracted, so that it should be possible to achieve an
effective description through an appropriately rescaled action. While
initially only a classical rescaling was
considered~\cite{Verlinde,Arefeva1,Arefeva2,Orland:2008vg}, in recent
years 
the effects of quantum corrections have been computed by means of
anisotropic renormalisation in the continuum
theory~\cite{Orland,Cubero:2011ut,Cubero:2012nw}. Here we will
consider the same problem in the lattice approach, which will allow us
to clarify, to some extent, the structure of the action in the limit
of large anisotropy. 
Notice that the anisotropy class (2+2) is the same considered in
Ref.~\cite{Bur}.  

On the lattice, the tree-level anisotropic action is given by
Eq.~\eqref{eq:tree_action} with the following anisotropy parameters, 
\begin{equation}
  \label{eq:LR_anis}
\lambda_4^{(LR)}(\xi)=\lambda_1^{(LR)}(\xi)=\xi\,,\qquad
\lambda_2^{(LR)}(\xi)=\lambda_3^{(LR)}(\xi)=1\,.  
\end{equation}
In the following, the superscript $LR$ is used to specify that this
particular choice has been made. 
We will refer to directions 4 and 1 as
longitudinal, and directions 2 and 3 as transverse, and use the
notation $n_\parallel=(n_4,n_1)$, $n_\perp=(n_2,n_3)$,
$a_\parallel=a_4=a_1=a/\xi$, $a_\perp=a_2=a_3=a$.
The plaquette coefficients $C^{(LR)}_{\mu\nu}$ in the anisotropic
action read  
\begin{equation}
  \label{eq:orland_coeff}
 C^{(LR)}_{23}(\xi)=\f{1}{\xi^2}\,,\quad
C^{(LR)}_{24}(\xi)=C^{(LR)}_{21}(\xi)=C^{(LR)}_{34}(\xi)=C^{(LR)}_{31}(\xi)=1\,,\quad   
  C^{(LR)}_{41}(\xi)=\xi^2\,. 
\end{equation}
The interesting case is that of large $\xi$. 
Taking na\"ively the  limit $\xi\to\infty$ in the tree-level action, the
transverse-transverse plaquette term drops from the action, while the
longitudinal-longitudinal term yields essentially a ``delta
function'' forcing the longitudinal links to be trivial. 
The resulting effective action would read
\begin{equation}
  \label{eq:2D_orland_tree}
  \begin{aligned} S_{\rm lat}^{\rm tree} &\mathop\to_{\xi\to\infty}
    S^{(2D)} = \sum_{n_\perp} 
S_\chi^{(2)}(n_\perp)+S_\chi^{(3)}(n_\perp)\,, \\
S_\chi^{(\mu)}(n_\perp) &=
\f{\beta}{2N_c}\sum_{n_\parallel}\sum_{\alpha=4,1} \tr\{[\Delta^+_\alpha
U_\mu(n)][\Delta^+_\alpha U_\mu(n)]^\dag\}\,, 
  \end{aligned}
\end{equation}
which describes a set of independent 2D principal chiral models
involving the transverse link variables, each one
living in the longitudinal plane at a given point $n_\perp$ in the
transverse plane. 
Here $\Delta_\alpha^+$ has been redefined by omitting the
$\lambda_\alpha$ factor [see Eq.~\eqref{eq:cov_der}]. 
Taking into account quantum corrections,
however, a different coupling has to be used for each of the three
different kinds of plaquette terms, namely 
$\beta^{(LR)}_{\parallel\parallel}= \beta^{(LR)}_{41}$ for the
longitudinal-longitudinal term,
$\beta^{(LR)}_{\perp\perp}=\beta^{(LR)}_{23}$ for the
transverse-transverse term
and $\beta^{(LR)}_{\parallel\perp}=
\beta^{(LR)}_{42}=\beta^{(LR)}_{43}=\beta^{(LR)}_{12}=\beta^{(LR)}_{13}$
for the longitudinal-transverse terms.
Recall that the
quantum corrections are of the form
\begin{equation}
  \label{eq:largeT7}
  \begin{aligned}
    {\cal K}^{(LR)}_{\mu\nu}(\xi) 
&= 
    \frac{11}{3}\frac{N_c}{(4\pi)^2}\left[-\gamma + 
    \frac{64}{33}\right] + \Delta {\cal G}^{(LR)}_{\mu\nu}(\xi) +
\Delta {\cal Z}^{(LR)}_{\mu\nu}(\xi) \\ &= -\beta_0 \log c^2 + \Delta
{\cal G}^{(LR)}_{\mu\nu}(\xi) + 
\Delta {\cal Z}^{(LR)}_{\mu\nu}(\xi)\,,
  \end{aligned}
\end{equation}
with $\Delta {\cal G}^{(LR)}_{\mu\nu}$ and $\Delta {\cal Z}^{(LR)}_{\mu\nu}$
containing respectively the contributions of the ${\cal G}$- and
${\cal Z}$-integrals, Eqs.~\eqref{eq:integrals} and
\eqref{eq:integrals2}. Obviously, $\Delta {\cal G}^{(LR)}_{42}=
\Delta {\cal G}^{(LR)}_{43}=
\Delta {\cal G}^{(LR)}_{12}=
\Delta {\cal G}^{(LR)}_{13}$, and similarly for
$\Delta {\cal Z}^{(LR)}_{\mu\nu}$. 
Using the large-$\xi$ behaviour of
these integrals, derived in Appendix \ref{app:GZanal}, one gets 
\begin{equation}
  \label{eq:orland_beta}
  \begin{aligned}
    \f{\beta^{(LR)}_{\perp\perp}(a,\xi)}{2N_c} &=  
     \beta_0  \log\f{1}{(a\Lambda c)^2}  
    + 2 \f{N_c^2-1}{2N_c}\f{1}{4\pi}\log\xi^2 
+ \Delta {\cal G}_{\perp\perp}^{(LR),fin}(\xi) 
+\Delta {\cal  Z}_{\perp\perp}^{(LR),fin}(\xi)\,,\\ 
\f{\beta^{(LR)}_{\parallel\parallel}(a,\xi)}{2N_c} &=  
     \beta_0  \log\f{1}{(a\Lambda c)^2}  
    + \Delta {\cal G}_{\parallel\parallel}^{(LR),fin}(\xi) +
\Delta {\cal Z}_{\parallel\parallel}^{(LR),fin}(\xi)\,,\\ 
\f{\beta^{(LR)}_{\parallel\perp}(a,\xi)}{2N_c} &=  
     \beta_0  \log\f{1}{(a\Lambda c)^2}  +
\f{N_c}{4}\f{1}{4\pi}\log\xi^2
    + \Delta {\cal G}_{\parallel\perp}^{(LR),fin}(\xi) 
+\Delta {\cal Z}_{\parallel\perp}^{(LR),fin}(\xi)\,,
  \end{aligned}
\end{equation}
where the superscript $fin$ on a quantity indicates that it 
is finite in the limit $\xi\to\infty$. 
If we keep the transverse spacing $a_\perp=a$ fixed, then taking
$\xi\to\infty$ means taking $a_\parallel = a_\perp/\xi$ to zero, i.e.,
taking the continuum limit in the longitudinal plane only. One then
sees that in general it is not allowed to discard the
transverse-transverse plaquette term, since  
$\sum_{n_\parallel} \f{\beta_{\perp\perp}}{2N_c}\xi^{-2} {\cal
  P}_{23} = \sum_{n_\parallel} \f{\beta_{\perp\perp}}{2N_c}
(\f{a_\parallel}{a_\perp})^2{\cal P}_{23}$ 
contains the right power of $a_\parallel$ to become the
two-dimensional integral over the longitudinal plane in the limit
$a_\parallel\to 0$. 

The action can now be recast in a form appropriate for a set of
coupled two-dimensional principal chiral models. To this end, 
it is convenient to introduce the following couplings, 
\begin{equation}
  \label{eq:2D_beta}
  \begin{aligned}
\beta_{(2D)}(a_\parallel,a_\perp) & = 
    \f{\beta^{(LR)}_{\parallel\perp}(a,\xi)}{2N_c}\bigg|_{\Oc(\xi^0)}
= \f{N_c}{2}
\f{1}{4\pi}\log\f{1}{a_\parallel\Lambda^{(2D)}(a_\perp)}\,,\\
\tilde\beta_{(2D)}(a_\parallel,a_\perp) & = 
\f{\beta^{(LR)}_{\perp\perp}(a,\xi)}{2N_c}\bigg|_{\Oc(\xi^0)}
= \f{N_c^2-1}{2N_c}
\f{1}{\pi}\log\f{1}{a_\parallel\tilde\Lambda^{(2D)}(a_\perp)} \,,    \\
\hat\beta_{(2D)}(a_\perp) &= 
    \f{\beta^{(LR)}_{\parallel\parallel}(a,\xi)}{2N_c}\bigg|_{\Oc(\xi^0)} =
     \beta_0  \log\f{1}{(a_\perp \Lambda c )^2}
 \,,
  \end{aligned}
\end{equation}
where the $a_\perp$-dependent scales $\Lambda^{(2D)}$ and
$\tilde\Lambda^{(2D)}$ are given in terms of the original 
$\Lambda$-scale as follows, 
\begin{equation}
  \label{eq:lambda2d}
  \begin{aligned}
    \Lambda^{(2D)}(a_\perp) &= 
\Lambda c(a_\perp\Lambda c)^{\f{16\pi\beta_0}{N_c}-1}
e^{-\f{8\pi}{N_c}[\Delta {\cal
        G}_{\parallel\perp}^{(LR),fin}(\infty) + \Delta {\cal
        Z}_{\parallel\perp}^{(LR),fin}(\infty)] }\,,\\
    \tilde\Lambda^{(2D)}(a_\perp) &=
\Lambda c(a_\perp\Lambda c)^{\f{4\pi\beta_0 N_c}{N_c^2-1}-1}
e^{-\f{2\pi N_c}{N_c^2-1}[\Delta {\cal
        G}_{\perp\perp}^{(LR),fin}(\infty) + \Delta {\cal
        Z}_{\perp\perp}^{(LR),fin}(\infty)] }\,. 
  \end{aligned}
\end{equation} 
The action can be equivalently written as follows,
\begin{equation}
  \label{eq:2D_orland_1loop_0}
S_{\rm lat} = \sum_{n_\perp}
S_\chi^{(2)}(n_\perp)+S_\chi^{(3)}(n_\perp) + S_{\rm int1}(n_\perp) +
S_{\rm int2}(n_\perp) \,,  
\end{equation}
where $S_\chi^{(\mu)}$ correspond to principal chiral models,
\begin{equation}
  \label{eq:2D_orland_1loop_1}
S_\chi^{(\mu)}(n_\perp) =
\beta_{(2D)}(a_\parallel,a_\perp)\sum_{n_\parallel}\sum_{\alpha=4,1} \tr\{[\Delta^+_\alpha
U_\mu(n)][\Delta^+_\alpha U_\mu(n)]^\dag\}\,,   
\end{equation}
and the interaction terms read
\begin{equation}
  \label{eq:2D_orland_1loop}
  \begin{aligned} 
S_{\rm int1}(n_\perp) &=
\tilde\beta_{(2D)}(a_\parallel,a_\perp)\sum_{n_\parallel}\f{a_\parallel^2}{a_\perp^2}{\cal
  P}_{23}(n)\,,\\
S_{\rm int2}(n_\perp) &=
\beta_{(2D)}(a_\parallel,a_\perp)\sum_{\mu=2,3}
\sum_{n_\parallel}\sum_{\alpha=4,1} \left[2N_c {\cal
    P}_{\mu\alpha}(n) - \tr\{[\Delta^+_\alpha
U_\mu(n)][\Delta^+_\alpha U_\mu(n)]^\dag\}
\right] \\ & \phantom{=}
+\hat\beta_{(2D)}(a_\perp)
\sum_{n_\parallel}2N_c\f{a_\perp^2}{a_\parallel^2}{\cal P}_{41}(n)\,. 
  \end{aligned}
\end{equation}
The only approximation made here is to discard $o(\xi^0)$ terms in the
couplings, so that this is just a rewriting of the original action in
the limit of large $\xi$. Nevertheless, this expression displays a 
remarkable feature: the coupling $\beta_{(2D)}$ is precisely the one
appropriate for a 2D principal chiral model with 
lattice spacing $a_\parallel$, to one-loop accuracy (see, e.g.,
Ref.~\cite{Polyakov}). The principal chiral models are clearly not 
independent, with the precise 
form of the interaction dictated by the full 4D action. Notice
that identifying the longitudinal links with $U_\mu =
\exp\{ia_\parallel \Q_\mu\}$ and expanding in powers of $a_\parallel$,
the summands in the interaction term $S_{\rm int2}$ are of order
$\Oc(a_\parallel^2)$, as appropriate to obtain an integral over 
the longitudinal plane in the na\"ive $a_\parallel\to 0$ limit, so
there is no reason to discard these contributions.\footnote{We 
  notice that working in the axial gauge $U_1=1$ and expanding $S_{\rm
    int2}$ to $\Oc(a_\parallel^2)$, the resulting expression is
  quadratic in $\Q_4$ and the corresponding integration can be carried
  out. This leads to the appearence of complicated, non-local
  interaction terms involving the transverse link variables.}  
It is not surprising that the interaction terms cannot be be
neglected {\it a priori}: after all, no matter how anisotropic the
lattice is made, by construction the action has to describe QCD in the
continuum limit. The possibility or not to neglect the interaction
terms will depend on the properties of the specific observables
relevant to the study of high-energy processes. 

The expectation values of the different plaquette terms can be used to
estimate the range of applicability of the expressions above.
Using Eq.~\eqref{eq:plaq_2} one gets to lowest order [see
Eq.~\eqref{eq:Zlargexi}] 
\begin{equation}
  \label{eq:plaq_long}
  \begin{aligned}
   \la {\cal P}_{41}\ra &= g^2\f{N_c^2-1}{N_c}\f{{\cal
          Z}^{(LR)}_\parallel(\xi)}{\xi^2} \simeq
g^2\f{N_c^2-1}{N_c}\f{z_{10}}{\xi^2}
\,,\\
\la    {\cal P}_{42}\ra &= g^2\f{N_c^2-1}{2N_c}\left(
      {\cal Z}^{(LR)}_\parallel(\xi) + \f{{\cal
          Z}^{(LR)}_\perp(\xi)}{\xi^2}  \right) \simeq 
g^2\f{N_c^2-1}{2N_c}z_{10}
\,,\\
  \la  {\cal P}_{23}\ra &= g^2\f{N_c^2-1}{N_c}{\cal
          Z}^{(LR)}_\perp(\xi) \simeq
g^2\f{N_c^2-1}{N_c}\f{1}{4\pi}\log\xi^2
\,,
  \end{aligned}
\end{equation}
with $z_{10}$ a constant defined in Eq.~\eqref{eq:aux_z}, 
so that in order to have small fluctuations one needs $g^2\log\xi\ll
1$. Together with the basic assumption $g^2\ll 1$, and the fact that
we work here at $\xi \gg 1$, the requirement $\la  {\cal
  P}_{23}\ra\ll 1$ defines the range of applicability of
perturbation theory, which in terms of the lattice spacings reads
\begin{equation}
  \label{eq:PT_range}
  1 \gg a_\perp\Lambda  
  \gg a_\parallel \Lambda  
  \gg (a_\perp \Lambda )^{1+\f{4\pi\beta_0 N_c}{N_c^2-1}}
\ge (a_\perp \Lambda )^2\,.
\end{equation}
The important fact is that Eq.~\eqref{eq:PT_range} does not allow us to
strictly take the continuum limit in the longitudinal plane before
taking $a_\perp$ to zero. This was already suggested in
Ref.~\cite{Orland}, although there it is claimed that perturbation
theory makes sense only for $\xi$ slightly larger than 1; 
according to our results, a much larger region seems to be accessible. 

A comparison of our results with those of
Refs.~\cite{Orland,Cubero:2011ut,Cubero:2012nw} is not
straightforward. First of all, since we use a different
regularisation, we expect different finite contributions to
the renormalisation of the couplings in the limit $a\to 0$ (at fixed
$\xi$); ultraviolet divergences, on the other hand, have to be the
same. Indeed, to account for a change in the cutoff, Orland and
collaborators integrate over an anisotropic ellipsoidal shell in
momentum space, while on the lattice a change in the cutoff 
requires us to integrate over an anisotropic parallelepipedal shell.
It would be interesting to compare the divergent terms in the limit
$\xi\to\infty$, but in
Refs.~\cite{Orland,Cubero:2011ut,Cubero:2012nw} only the case
$\xi\gtrsim 1$ is studied.  

We conclude by noticing that a similar recasting of the action can
be done also in the case discussed in Section \ref{sec:AC},
considering the limit of large $\barT$. The results are briefly
discussed in Appendix \ref{app:largeTaction}.

\secspace
\section{Conclusions}
\label{sec:concl}

In this paper we have performed the renormalisation of 
$SU(N_c)$ gauge theories on a general four-dimensional anisotropic 
lattice, with different lattice spacings in the four directions,
using perturbation theory to one-loop order and the
background field method on the lattice (Section
\ref{sec:anis_ren}). 
To avoid the complications related to the introduction of fermions on the
lattice, we have discussed here the pure-gauge case only. 
For general anisotropy, the various couplings in the gauge action
need to be properly tuned in order to recover $O(4)$ invariance in the
continuum limit, as already observed in Ref.~\cite{Bur}. In practice,
however, only two parameters need to be tuned for this purpose, which
reduce to one if there is at least a pair of equal lattice spacings
(and to none in the 3+1 case). A simple nonperturbative scheme for
this tuning, based on the string tensions obtained in different
lattice planes, has also been proposed.  

In Section \ref{sec:AC}, the possibility to vary continuously the
anisotropy parameters has been exploited in the context of the
nonperturbative approach to soft high-energy hadron-hadron scattering
based on Wilson loops~\cite{Nachtmann91,DFK,Nachtmann97,BN,Dosch,LLCM1,
pomeron-book,reggeon}, in order to refine the arguments of
Ref.~\cite{GM2009} on the analyticity properties of the relevant
Wilson-loop correlators. The results reported here give further
support to the possibility of performing the desired analytic
continuation between Euclidean and Minkowski space, and thus on the
very possibility of using Euclidean techniques to study soft
high-energy processes. This is particularly important in the light of
recent progress on the problem of hadronic total cross
sections~\cite{sigtot,sigtot_comments}, which is based on the
possibility of recovering the physical amplitudes starting from 
Euclidean space. 

In Section \ref{sec:2dmodel} we have applied our results to the
longitudinally rescaled actions considered in Refs.~\cite{Verlinde,
Arefeva1,Arefeva2,Orland:2008vg,Orland,Cubero:2011ut,Cubero:2012nw} to 
study high-energy scattering in QCD. At the classical level, in the
limit of large anisotropy the action reduces to that of a set of
coupled two-dimensional principal chiral models, living in the
longitudinal plane at each point of the transverse plane. Our main
result in this context is that this interpretation holds also at the
one-loop level, as the bare coupling resulting in the free part of
each principal chiral model behaves appropriately as a function 
of the longitudinal lattice spacing. The precise form of the
interactions among the principal chiral models is dictated by the full
gauge action. However, the limit of large anisotropy cannot be taken
independently of the continuum limit, at least in the perturbative
approach. Indeed, the requirement of small gauge field fluctuations
defines a range of validity of the form $1\gg a_\perp\Lambda \gg
a_\parallel\Lambda \gg (a_\perp \Lambda)^{1+\gamma}$ for the longitudinal and
transverse lattice spacings $a_\parallel$ and $a_\perp$, where
$\Lambda$ is the QCD scale and $\gamma>0$. Nevertheless, our findings
suggest that there may be a deeper relation between gauge theories and
principal chiral models than just at the classical level. 

There are several open directions for future studies. An obvious
possibility is the inclusion of fermions in the analysis. 
This is particularly relevant to the nonperturbative approach to soft
high-energy scattering, since the presence or not of dynamical
fermions seems to have large effects on total cross
sections~\cite{sigtot,sigtot_comments}. It would be interesting to
extend the perturbative analysis to non-orthogonal lattices, which
would allow us to use on-axis Wilson loops in the relevant lattice
correlator. However, in this case more terms appear in the action, so
leading to a more intricate calculation.

\secspace
\section*{Acknowledgments}
I want to thank E.~Meggiolaro and T.~G.~Kov\'acs for many useful
remarks. I also acknowledge interesting discussions and correspondence
with G.~Burgio, E.~Follana, S.~Katz, and P.~Orland.
This work is supported by the Hungarian Academy of Sciences under
``Lend\"ulet'' grant No. LP2011-011.

\newpage
\appendix

\section{The ${\cal G}$- and ${\cal Z}$-integrals}
\label{app:GZanal}

In the expression for the one-loop contributions ${\cal K}_{\mu\nu}$,
Eq.~\eqref{eq:K_app}, there appear a few 
integrals involving the modif\mbox{}ied Bessel functions 
of the f\mbox{}irst kind $I_n(z)$, which are special cases of the
integrals 
\begin{equation}
  \label{eq:integrals}
  \begin{aligned}
    {\cal G}_n(\lambda) &=
\left\{
  \begin{aligned}
   & \int_0^\infty d\rho\, \rho \left[\prod_{\alpha=1}^4
    \lambda_\alpha 
\modbes_0^{\,(n_\alpha)}(2\lambda_{\alpha}^2\rho)\right]\,,\quad
  n \neq (0,0,0,0)\,,
\\
  &  \int_0^\infty d\rho\, \rho \left\{\left[\prod_{\alpha=1}^4\lambda_\alpha
    \modbes_0(2\lambda_{\alpha}^2\rho)\right]
    -\Theta(\rho-1)\frac{1}{(4\pi\rho)^2}\right\}\,,\quad
  n = (0,0,0,0)\,,
  \end{aligned}\right. \\
    {\cal Z}_n(\lambda) &=  \int_0^\infty d\rho\, \left[\prod_{\alpha=1}^4
    \lambda_\alpha 
\modbes_0^{\,(n_\alpha)}(2\lambda_{\alpha}^2\rho)
\right]\,,
  \end{aligned}
\end{equation}
defined for a general four-vector of integers $n$, where
\begin{equation}
  \label{eq:mod_bes}
  \modbes_n(z)
\equiv e^{-z} I_n(z)\,, \qquad \modbes_n^{\,(m)}(z)\equiv(-\de/\de
z)^{m}\modbes_n(z)\,,
\end{equation}
and $\Theta(z)$ is the step function. In
particular, in Eq.~\eqref{eq:K_app} we have denoted as follows the
relevant cases,
\begin{equation}
  \label{eq:integrals2}
  \begin{aligned}
    & {\cal G}_{\mu\nu} 
    = {\cal G}_n \big |_{n_\alpha =
      \delta_{\alpha\mu}+\delta_{\alpha\nu}} \,, &&& 
&    {\cal G}_{\mu} 
    = {\cal G}_n \big |_{n_\alpha =
      \delta_{\alpha\mu}}\,, &&& 
{\cal G} = {\cal G}_n |_{n_\alpha = 0} \,, \\
&    {\cal Z}_{\mu} 
    = {\cal Z}_n \big |_{n_\alpha =
      \delta_{\alpha\mu}}\,, &&& 
 &   {\cal Z}={\cal Z}_n |_{n_\alpha = 0}\,.
  \end{aligned}
\end{equation}
These integrals are not all independent; in particular, the following
sum rules hold, 
\begin{equation}
  \label{eq:sum_rules}
  \begin{aligned}
    & \sum_{\mu=1}^4 2\lambda_\mu^2 {\cal Z}_\mu=
    \prod_{\mu=1}^4\lambda_\mu = {\cal J}^{-1}\,, &&&
    & \sum_{\mu=1}^4 2\lambda_\mu^2 {\cal G}_\mu = {\cal Z}\,,\\ 
    & 2\lambda_\mu^2 {\cal G}_\mu + \sum_{\nu\neq \mu} 2\lambda_\nu^2
    {\cal G}_{\mu\nu} = \sum_{\nu\neq \mu} 2\lambda_\nu^2  {\cal
      G}_\nu  \,.
  \end{aligned}
\end{equation}
A global rescaling $\lambda_\alpha\to\zeta\lambda_\alpha$ ($\zeta>0$)
of the anisotropy parameters can be essentially reabsorbed in ${\cal
  Z}_n$ and in ${\cal G}_n$ by changing variables to
$\rho'=\zeta^2\rho$, which brings about a multiplicative 
factor for ${\cal Z}_n$, and an additive contribution proportional to
$\log\zeta$ to ${\cal G}_0$. More precisely, we find
\begin{equation}
  \label{eq:rescale_app}
  {\cal Z}_n(\zeta\lambda) =  \zeta^2{\cal Z}_n(\lambda) \,,\qquad
  {\cal G}_n(\zeta\lambda) = \left\{
    \begin{aligned}
      & {\cal G}_n(\lambda)      \,, \quad n\neq 0\,,\\
      & {\cal G}_0(\lambda) + \f{1}{(4\pi)^2}\log\zeta^2
        \,, \quad n= 0\,.
    \end{aligned}\right.
\end{equation}

\subsection{Analyticity properties}

We discuss now the analyticity properties of ${\cal K}_{\mu\nu}$. It
is clear that for $\lambda_\alpha\neq 0 \,\forall\alpha$ these depend only on
the analyticity properties of the ${\cal G}$- and ${\cal
  Z}$-integrals defined in Eq.~\eqref{eq:sum_rules}. Since these are
integrals of analytic functions of $\rho$ and $\{\lambda_\alpha\}$, it
suffices to show that they converge uniformly in $\{\lambda_\alpha\}$
within some complex domain. In turn, a sufficient condition
for this is that we can bound the modulus of the integrand uniformly
in $\{\lambda_\alpha\}$ by some function $f$, whose integral is also
convergent. To do this, we need the following inequalities,
\begin{equation}
  \label{eq:ineq}
  \begin{aligned}
      |\modbes_0(z)| &\le  \modbes_0(\Re z)\,,\quad
      |\modbes_1(z)| \le  \modbes_0(\Re z)\,,\quad
      |\modbes_0(z)| \le 1 \quad \text{if}~~\Re z\ge 0\,,
  \end{aligned}
\end{equation}
which are easily proved using the integral representation for
$I_n(z)$. We also need 
the monotonicity property
\begin{equation}
  \label{eq:ineq2}
    \f{\de}{\de x} \modbes_0(x)\le 0\,,\quad \forall\,x\in\mathbb{R}\,,
\end{equation}
and the asymptotic behaviour of $\modbes_0(z)$,
\begin{equation}
  \label{eq:asymptoticI}
    \modbes_0(z) \sim \f{1}{\sqrt{2\pi z}}\left( 1 
- \f{1}{8z} +
    \Oc(z^{-2})\right)\,, 
\end{equation}
valid for $|\arg z| <  \pi$ (see, e.g., Ref.~\cite{GradRhiz}). 
\paragraph{1} The quantities ${\cal Z}$ and ${\cal Z}_\mu$ are given
by the product of the analytic factor ${\cal
  J}^{-1}=\prod_\alpha\lambda_\alpha$ and an integral of the product
of functions $\modbes_0$ and, in the case of ${\cal Z}_\mu$, also
$\modbes_0-\modbes_1$, so that we may write
\begin{equation}
  \label{eq:Z2}
  {\cal Z}(\{\lambda_\alpha\}) = {\cal J}^{-1}\tilde{\cal
    Z}(\{\lambda_\alpha^2\})\,,\quad
  {\cal Z}_\mu(\{\lambda_\alpha\}) = {\cal J}^{-1}\tilde{\cal
    Z}_\mu(\{\lambda_\alpha^2\})\,. 
\end{equation}
For $\{\lambda_\alpha\}$ such that for every
$\alpha$ one has $\Re\lambda_\alpha^2\in [u_\alpha,v_\alpha]$, with
$u_\alpha,v_\alpha\in\mathbb{R}$, $0<u_\alpha<v_\alpha<\infty$, the
first two inequalities in Eq.~\eqref{eq:ineq} and the monotonicity
property Eq.~\eqref{eq:ineq2} tell us that a possible choice for
$f(\rho)$ to bound the modulus of the integrands both in $\tilde{\cal
  Z}$ and $\tilde{\cal Z}_\mu$ is $f(\rho)=2\prod_\alpha f_\alpha(\rho)$,
$f_\alpha(\rho)=\modbes_0(2u_\alpha \rho)$. In particular, this shows
that $\tilde{\cal Z}$ and $\tilde{\cal Z}_\mu$ are analytic functions
of $\{\lambda_\alpha^2\}$.

\paragraph{2} To study ${\cal G}$ we split the integral into two parts,
$\int_0^\infty= \int_0^1 + \int_1^\infty$. For the first piece, the
third inequality in Eq.~\eqref{eq:ineq} indicates that we can take
$f(\rho)=\rho$. The integrand of the second piece is conveniently
written as
\begin{equation}
  \label{eq:Cfunc}
      \rho\left( \prod_{\alpha=1}^4
    \lambda_\alpha\modbes_0(2\lambda_\alpha^2 \rho) -\f{1}{(4\pi \rho)^2
    }\right)
= \f{1}{(4\pi \rho)^2} \tilde f(\{\lambda_\alpha\},\rho)\,,
\end{equation}
where $\tilde f$ is analytic $\forall\,\lambda$ and $\rho\neq 0$, and
furthermore it is certainly bounded for $\Re\lambda_\alpha^2\in
[u_\alpha,v_\alpha]$ and $\rho\in [0,
\infty)$, 
since it has a finite limit as $\rho\to\infty$, see
Eq.~\eqref{eq:asymptoticI}. In this case we can then take $f(\rho) =
M/(4\pi\rho)^2$ for a properly chosen constant $M$.

\paragraph{3} Finally, analyticity properties of ${\cal G}_\mu$ and
${\cal G}_{\mu\nu}$ are inherited from ${\cal Z}$ and ${\cal
  Z}_\mu$. Indeed, since one can bring derivatives under the sign of
integral due to uniform convergence, one shows immediately that
\begin{equation}
  \label{eq:gmgmn}
  \begin{aligned}
    \lambda_\mu \f{\de}{\de\lambda_\mu}{\cal Z}(\lambda) 
    &={\cal Z}(\lambda) - 4\lambda_\mu^2{\cal G}_\mu(\lambda) \,,\\
    \lambda_\nu \f{\de}{\de\lambda_\nu}{\cal Z}_\mu(\lambda) 
    &={\cal Z}_\mu(\lambda) - 4\lambda_\nu^2{\cal G}_{\mu\nu}(\lambda)
    \quad(\nu\neq\mu)\,.
  \end{aligned}
\end{equation}
Notice that ${\cal G}_{\mu}$ and ${\cal G}_{\mu\nu}$ are of the form
\begin{equation}
  \label{eq:G2}
  {\cal G}_\mu(\{\lambda_\alpha\}) = {\cal J}^{-1}\tilde{\cal
    G}_\mu(\{\lambda_\alpha^2\})\,,\quad
  {\cal G}_{\mu\nu}(\{\lambda_\alpha\}) = {\cal J}^{-1}\tilde{\cal
    G}_{\mu\nu}(\{\lambda_\alpha^2\})\,, 
\end{equation}
with  $\tilde{\cal G}_{\mu}$ and $\tilde{\cal G}_{\mu\nu}$ analytic in
$\{\lambda_\alpha^2\}$.

In conclusion, ${\cal K}_{\mu\nu}$ are analytic in any compact domain
with $\Re\lambda_\alpha^2 > 0$, $\forall \alpha$. For our purposes, it
is convenient to extend further the domain of analyticity. To this
end, notice that for real positive $\lambda_\alpha$, one can
rewrite ${\cal Z}$, ${\cal Z}_\mu$, ${\cal G}_\mu$ and ${\cal
  G}_{\mu\nu}$ as follows by exploiting their
behaviour under global rescaling, Eq.~\eqref{eq:rescale_app},
\begin{equation}
  \label{eq:rew_ZG}
  \begin{aligned}
    {\cal Z}(\{\lambda_\alpha\}) &= \tilde{\cal Z}(\{{\cal
      J}\lambda_\alpha^2\})\,, &&&
  {\cal Z}_\mu(\{\lambda_\alpha\}) &= \tilde{\cal Z}_\mu(\{{\cal
    J}\lambda_\alpha^2\})\,,\\
  {\cal G}_\mu(\{\lambda_\alpha\}) &= {\cal J}\tilde{\cal G}_\mu(\{{\cal
    J}\lambda_\alpha^2\})\,, &&&
  {\cal G}_{\mu\nu}(\{\lambda_\alpha\}) &= {\cal J}\tilde{\cal G}_{\mu\nu}(\{{\cal
    J}\lambda_\alpha^2\}) \,,
  \end{aligned}
\end{equation}
where Eqs.~\eqref{eq:Z2} and \eqref{eq:G2} have been used. The domain
of analyticity of these quantities can thus be straightforwardly
extended to $\Re({\cal J}\lambda_\alpha^2)>0$. Furthermore, for real
positive $\lambda_\alpha$, one has
\begin{equation}
  \label{eq:G_rev}
{\cal G} =  \int d\rho\, \rho \left[{\cal J} \left(\prod_\alpha
    \modbes_0(2{\cal J}\lambda_{\alpha}^2\rho)\right)
    -\Theta(\rho-1)\frac{1}{(4\pi\rho)^2}\right] -
  \frac{1}{(4\pi)^2}\log{\cal J}\,,
\end{equation}
where we have used Eq.~\eqref{eq:rescale_app} again. By the same token
used above in point 2, the first term in Eq.~\eqref{eq:G_rev} is
analytic for $\Re{\cal J}\lambda_\alpha^2>0$. The logarithmic term is an
analytic function in the cut complex plane for $|\arg{\cal J}|<\pi$,
so we conclude that ${\cal K}_{\mu\nu}$ are analytic also in the
domain defined by $\Re({\cal J}\lambda_\alpha^2)>0$,  $|\arg{\cal
  J}|<\pi$.  

We now analyse the specific case discussed in Section \ref{sec:AC},
corresponding to the following choice of anisotropy parameters,
\begin{equation}
  \label{eq:anis_param_app}
  \lambda_4(\theta,\barT) = \frac{1}{\sqrt{2}\barT \cos\frac{\theta}{2} } \,,
\quad
  \lambda_1(\theta,\barT) = \frac{1}{\sqrt{2}\barT \sin\frac{\theta}{2}} \,,
\quad 
  \lambda_2(\theta,\barT) = \lambda_3(\theta,\barT) = 1\,,
\end{equation}
which, in the light of 
the extension of the analyticity domain discussed above,
can be recast more conveniently as follows,
\begin{equation}
  \label{eq:relations_app2}
  \begin{aligned}
&    {\cal J}(\theta,\barT)\lambda_4^2(\theta,\barT)= 
    {C_{42}(\theta,\barT)}\,, 
    \quad
    {\cal J}(\theta,\barT)  \lambda_1^2(\theta,\barT)=
    C_{12}(\theta,\barT)\,, \\
 &   {\cal J}(\theta,\barT) = {\cal J}(\theta,\barT)
    \lambda_2^2(\theta,\barT)={\cal 
      J}(\theta,\barT)  \lambda_3^2(\theta,\barT)={C}_{23}(\theta,\barT) \,. 
  \end{aligned}
\end{equation}
As functions of complex angle and length, $C_{\mu\nu}(w,\barT)$ are
analytic everywhere, except at $w=n\pi$ with $n\in \mathbb{Z}$, and
$|\barT|=0$. Since the domain ${\cal D}$ considered in Section
\ref{sec:AC} is defined by $\Re C_{\mu\nu}(w,\barT)>0$, in ${\cal D}$
one has that $|\arg{\cal J}(w,\barT)|<\f{\pi}{2}$ and $\Re({\cal
  J}(w,\barT)\lambda_\alpha^2(w,\barT))>0$, so 
that the ${\cal G}$- and ${\cal Z}$-integrals are analytic there, and 
in conclusion the one-loop corrections ${\cal
  K}_{\mu\nu}(w,\barT)$ are analytic in ${\cal D}$. 

\subsection{Large-$\barT$ behaviour}

We now determine, for real $\barT$, the large-$\barT$ behaviour of the
${\cal Z}$- and ${\cal G}$-integrals for the choice of anisotropy
parameters of Eqs.~\eqref{eq:anis_param} and
\eqref{eq:anis_param_app}. To this end, it is  
convenient to define the following auxiliary quantities, 
\begin{equation}
  \label{eq:largeT1}
  \begin{aligned}
    B_n(\theta,\barT) &= \int_0^\infty d\rho\, \left[\prod_\alpha
      \lambda_\alpha \modbes_{n_\alpha}(2\lambda_{\alpha}^2
      \rho)\right]\,, \\
    D_n(\theta,\barT) &= \int_0^\infty d\rho\, \rho\left[\prod_\alpha
      \lambda_\alpha \modbes_{n_\alpha}(2\lambda_{\alpha}^2
      \rho)
    -\Theta(\rho-1)\frac{1}{(4\pi\rho)^2}\right]
\,,
  \end{aligned}
\end{equation}
where $\{\lambda_\alpha\}$ are chosen according to
Eq.~\eqref{eq:anis_param}. It is straightforward to show that
\begin{equation}
  \label{eq:ZGfromBD}
  \begin{aligned}
    {\cal Z}&=B_n |_{n_\alpha = 0}\,,\quad {\cal Z}_\mu=B_n
    |_{n_\alpha = 0} - B_n |_{n_\alpha =
    \delta_{\alpha\mu}}\,,     \\
  {\cal G}&=D_n |_{n_\alpha = 0}\,,\quad {\cal G}_\mu = D_n
  |_{n_\alpha = 0}- D_n |_{n_\alpha = \delta_{\alpha\mu}}\,,  \\
 {\cal G}_{\mu\nu} &= D_n |_{n_\alpha = \delta_{\alpha\mu}+\delta_{\alpha\nu}} + D_n
  |_{n_\alpha = 0}- D_n |_{n_\alpha = \delta_{\alpha\mu}} - D_n
  |_{n_\alpha = \delta_{\alpha\nu}}\,.   
  \end{aligned}
\end{equation}
A rather simple calculation shows that at
large $\barT$
\begin{equation}
  \label{eq:largeT2}
  \begin{aligned}
    B_n(\theta,\barT) =&~ \f{1}{\barT^2\sin\theta}\Bigg\{
      \f{1}{4\pi}\modbes_{n_4}(0)\modbes_{n_1}(0)\log\barT^2 
 + b_n(\theta) + o(\barT^0)\Bigg\}\,,
\\ 
b_n(\theta) =&~ \modbes_{n_4}(0)\modbes_{n_1}(0) 
\int_0^1 d\rho\,\modbes_{n_2}(2\rho)
      \modbes_{n_3}(2\rho)\\ &+
 \f{1}{4\pi}\int_0^\infty\f{d\rho}{\rho}\,\left[
      \modbes_{n_4}\left(\f{\rho}{\cos^2\f{\theta}{2}}\right)
\modbes_{n_1}\left(\f{\rho}{\sin^2\f{\theta}{2}}\right)
- \Theta(1-\rho)\modbes_{n_4}(0)\modbes_{n_1}(0)\right]\,,\\
D_n(\theta,\barT) =&~ -\f{1}{(4\pi)^2}\log\barT^2  + d_n(\theta) +
o(\barT^0)\,,\\ 
d_n(\theta) =&~ \f{1}{4\pi}\int_0^\infty d\rho\,\left[
\f{1}{\sin\theta}
\modbes_{n_4}\left(\f{\rho}{\cos^2\f{\theta}{2}}\right) 
\modbes_{n_1}\left(\f{\rho}{\sin^2\f{\theta}{2}}\right)
- \Theta(\rho-1)\frac{1}{4\pi\rho}\right]\,.
  \end{aligned}
\end{equation}
It is now straightforward to obtain the large-$\barT$ behaviour of the
relevant quantities. For the ${\cal Z}$-integrals we have
\begin{equation}
  \label{eq:largeT3}
  \begin{aligned}
    {\cal Z}(\theta,\barT) =&~\f{1}{\barT^2\sin\theta}\Bigg\{
      \f{1}{4\pi}\log\barT^2 + z_{00} + \tilde z_{00}(\theta)
  + o(\barT^0)\Bigg\}\,,\\
    {\cal Z}_4(\theta,\barT) =&~\f{1}{\barT^2\sin\theta}\Bigg\{
      \f{1}{4\pi}\log\barT^2 + z_{00} + \tilde z_{10}(\theta)
  + o(\barT^0)\Bigg\}\,,\\
    {\cal Z}_1(\theta,\barT) =&~\f{1}{\barT^2\sin\theta}\Bigg\{
      \f{1}{4\pi}\log\barT^2 + z_{00} + \tilde z_{01}(\theta)
  + o(\barT^0)\Bigg\}\,,\\
    {\cal Z}_2(\theta,\barT) =&~ {\cal Z}_3(\theta,\barT)
    =\f{1}{\barT^2\sin\theta}\Bigg\{ z_{10}  
  + o(\barT^0)\Bigg\}\,,
  \end{aligned}
\end{equation}
where we have introduced the following quantities,
\begin{equation}
  \label{eq:aux_z}
  \begin{aligned}
z_{nm} =&~ \int_0^1{d\rho}\,
\modbes_{0}^{\,(n)}\left(2\rho\right) 
\modbes_{0}^{\,(m)}\left(2\rho\right) \,,
\\
\tilde z_{nm}(\theta) =&~ \f{1}{4\pi}\int_0^\infty\f{d\rho}{\rho}\,\left\{
\modbes_{0}^{\,(n)}\left(\f{\rho}{\cos^2\f{\theta}{2}}\right)
\modbes_{0}^{\,(m)}\left(\f{\rho}{\sin^2\f{\theta}{2}}\right)
- \modbes_{0}^{\,(n)}(0)\modbes_{0}^{\,(m)}(0)\Theta(1-\rho)\right\}\,.
  \end{aligned}
\end{equation}
For the ${\cal G}$-integrals we find
\begin{equation}
  \label{eq:largeT4}
  \begin{aligned}
{\cal G}(\theta,\barT) =&~ -\f{1}{(4\pi)^2}\log\barT^2  + \tilde
g_{00}(\theta) + o(\barT^0)\,, &&& {\cal G}_{41}(\theta,\barT) =&~  \tilde
g_{11}(\theta) + o(\barT^0)\,, \\
{\cal G}_4(\theta,\barT) =&~  \tilde
g_{10}(\theta) + o(\barT^0)\,, &&&
{\cal G}_1(\theta,\barT) =&~  \tilde
g_{01}(\theta) + o(\barT^0)\,,
  \end{aligned}
\end{equation}
where we have introduced the following quantities,
\begin{equation}
  \label{eq:largeT4bis}
  \begin{aligned}
\tilde{g}_{nm}(\theta) =&~ 
\f{1}{4\pi}\int_0^1d\rho\,
\f{1}{\sin\theta}
\modbes_{0}^{\,(n)}\left(\f{\rho}{\cos^2\f{\theta}{2}}\right)
\modbes_{0}^{\,(m)}\left(\f{\rho}{\sin^2\f{\theta}{2}}\right)
\,,\quad (n,m)\neq (0,0)\,,\\
\tilde{g}_{00}(\theta) =&~ 
\f{1}{4\pi}\int_0^\infty d\rho\,\left\{
\f{1}{\sin\theta}
\modbes_{0}\left(\f{\rho}{\cos^2\f{\theta}{2}}\right) 
\modbes_{0}\left(\f{\rho}{\sin^2\f{\theta}{2}}\right)
- \Theta(\rho-1)\frac{1}{4\pi\rho}\right\}\,,
  \end{aligned}
\end{equation}
while the remaining integrals are all $o(\barT^0)$. One can now easily
determine the contributions of both kinds of terms to ${\cal
  K}_{\mu\nu}$, namely
\begin{equation}
  \label{eq:deltaG}
  \begin{aligned}
    \Delta {\cal G}_{\mu\nu}(\theta,\barT)  &=  N_c\left[
      \frac{2}{3} {\cal G}_{\mu\nu}(\theta,\barT) -\frac{5}{3}\big( {\cal
        G}_{\mu}(\theta,\barT) + {\cal G}_{\nu}(\theta,\barT)\big) + 
      \frac{11}{3} {\cal G}(\theta,\barT)\right] \\ &=
    \beta_0  \log\f{1}{\barT^2} + \Delta {\cal G}_{\mu\nu}^{fin}(\theta,\barT)\,,
  \end{aligned}
  \end{equation}
where $\beta_0$ is defined in Eq.~\eqref{eq:beta_func}, and
\begin{equation}
  \label{eq:largeT5}
  \begin{aligned}
\Delta {\cal G}_{41}^{fin}(\theta,\barT) & = N_c\left[
\frac{11}{3}\tilde g_{00}(\theta)
-\frac{5}{3}\big(\tilde g_{10}(\theta)+\tilde g_{01}(\theta)\big) +\frac{2}{3}
  \tilde g_{11}(\theta)
\right]+ o(\barT^0)\,,\\
\Delta {\cal G}_{42}^{fin}(\theta,\barT) & = \Delta {\cal
  G}_{43}^{fin}(\theta,\barT) = N_c\left[ 
\frac{11}{3}\tilde g_{00}(\theta)
-\frac{5}{3}\tilde g_{10}(\theta)
\right] + o(\barT^0)\,,\\
\Delta {\cal G}_{12}^{fin}(\theta,\barT) & = \Delta {\cal
  G}_{13}^{fin}(\theta,\barT) = N_c\left[ 
\frac{11}{3}\tilde g_{00}(\theta)
-\frac{5}{3}\tilde g_{01}(\theta)
\right] + o(\barT^0)\,,\\
\Delta {\cal G}_{23}^{fin}(\theta,\barT) & = o(\barT^0)\,,
  \end{aligned}
\end{equation}
and
\begin{equation}
  \label{eq:deltaZ}
  \begin{aligned}
    \Delta {\cal Z}_{\mu\nu}(\theta,\barT) &=   \frac{N_c}{4}\left[
{\cal Z}(\theta,\barT)\left(\frac{1}{\lambda^2_\nu(\theta,\barT)} +
  \frac{1}{\lambda^2_\mu(\theta,\barT)} 
\right) -
\frac{{\cal Z}_\mu(\theta,\barT)}{\lambda^2_\nu(\theta,\barT)}
-\frac{{\cal Z}_\nu(\theta,\barT)}{\lambda^2_\mu(\theta,\barT)}
\right] \\ &~~
  + \frac{N_c^2-1}{2N_c}\left[
\frac{{\cal Z}_\mu(\theta,\barT)}{\lambda^2_\nu(\theta,\barT)}
    +\frac{{\cal Z}_\nu(\theta,\barT)}{\lambda^2_\mu(\theta,\barT)}
\right]\,,  
  \end{aligned}
\end{equation}
where $\Delta {\cal Z}_{\mu\nu}$ can be split into a divergent and a finite
part, 
\begin{equation}
  \label{eq:largeT6_0}
  \begin{aligned}
    \Delta {\cal Z}_{41}(\theta,\barT) &= \Delta {\cal
      Z}_{41}^{div}(\theta,\barT) + 
    \Delta {\cal Z}_{41}^{fin}(\theta,\barT)  \,,  \\  
\Delta {\cal Z}_{42}(\theta,\barT) &= \Delta {\cal Z}_{43}(\theta,\barT) =
\Delta {\cal Z}_{4\perp}^{div}(\theta,\barT) + \Delta {\cal
  Z}_{4\perp}^{fin}(\theta,\barT)  
\,,
\\
\Delta {\cal Z}_{12}(\theta,\barT) &= \Delta {\cal Z}_{13}(\theta,\barT) =
\Delta {\cal Z}_{1\perp}^{div}(\theta,\barT) + \Delta {\cal
  Z}_{1\perp}^{fin}(\theta,\barT) 
\,,  \\
\Delta {\cal Z}_{23}(\theta,\barT) &= \Delta {\cal
  Z}^{fin}_{23}(\theta,\barT) =o(\barT^0)\,,
  \end{aligned}
\end{equation}
with divergent parts given by
\begin{equation}
  \label{eq:largeT6_1}
  \begin{aligned}
\Delta {\cal Z}_{41}^{div}(\theta,\barT) &=
\frac{N_c^2-1}{2N_c}\f{2}{\sin\theta} 
  \f{1}{4\pi}\log\barT^2 \,,  \\
\Delta {\cal Z}_{4\perp}^{div}(\theta,\barT) &= 
 \cot\f{\theta}{2}\f{N_c}{4}
\f{1}{4\pi}\log\barT^2\,, 
&&& \Delta {\cal Z}_{1\perp}^{div}(\theta,\barT) &= 
 \tan\f{\theta}{2}\f{N_c}{4}
\f{1}{4\pi}\log\barT^2 \,,    
  \end{aligned}
\end{equation}
and finite parts given by
\begin{equation}
  \label{eq:largeT6}
  \begin{aligned}
\Delta {\cal Z}_{41}^{fin}(\theta,\barT) &=
\frac{N_c^2-1}{2N_c}\left[\f{2}{\sin\theta}z_{00} + 
  \cot\f{\theta}{2}\tilde z_{01}(\theta)  + \tan\f{\theta}{2}\tilde
  z_{10}(\theta)\right]
 & \\ &~~
+ \f{N_c}{4}\left[
\f{2}{\sin\theta}\tilde z_{00}(\theta) -  \cot\f{\theta}{2}\tilde
z_{01}(\theta)  - \tan\f{\theta}{2}\tilde   z_{10}(\theta)
\right] + o(\barT^0)\,,  \\
\Delta {\cal Z}_{4\perp}^{fin}(\theta,\barT) &= 
 \cot\f{\theta}{2}\left[\f{N_c}{4}\big(
 z_{00}+ \tilde z_{00}(\theta)\big) +
\frac{N_c^2-2}{4N_c}z_{10}\right]+ o(\barT^0)\,,
 \\
\Delta {\cal Z}_{1\perp}^{fin}(\theta,\barT) &= 
 \tan\f{\theta}{2}\left[\f{N_c}{4}\big(
 z_{00}+ \tilde z_{00}(\theta)\big) +
\frac{N_c^2-2}{4N_c}z_{10}\right]+ o(\barT^0)\,,
  \end{aligned}
\end{equation}
from which one can easily reconstruct
the behaviour of ${\cal K}_{\mu\nu}(\theta,\barT)$ up to $o(\barT^0)$.

The results above allow us to easily derive the large-$\xi$ behaviour of
the couplings when the anisotropy parameters are chosen appropriately
for the longitudinally rescaled action of Section \ref{sec:2dmodel},
i.e., $\lambda_4^{(LR)}=\lambda_1^{(LR)}=\xi$ and
$\lambda_2^{(LR)}=\lambda_3^{(LR)}=1$, see
Eq.~\eqref{eq:LR_anis}. This is accomplished through the following
steps. First of all, notice that $\lambda_\mu^{(LR)}$ are just a
particular case of $\lambda_\mu(\theta,\barT)$, namely
$\lambda_\mu^{(LR)}(\xi)=\lambda_\mu(\f{\pi}{2},\f{1}{\xi})$. Next, it
is straightforward to show that 
\begin{equation}
  \label{eq:BD_LR}
B_n({\textstyle\f{\pi}{2}},{\textstyle\f{1}{\xi}})=\xi^2
B_{\tilde n}({\textstyle\f{\pi}{2}},\xi)\,, \qquad
D_n({\textstyle\f{\pi}{2}},{\textstyle\f{1}{\xi}})=\f{1}{(4\pi)^2}\log\xi^2 +
D_{\tilde n}({\textstyle\f{\pi}{2}},\xi)\,,  
\end{equation}
where $\tilde n_\mu = n_{\tilde \mu}$ with
$\{\tilde\mu\}=\{\tilde 1,\tilde 2,\tilde 3,\tilde 4\}=\{3,4,1,2\}$.  
Finally, one easily shows that
\begin{equation}
  \label{eq:BD_LR_2}
\f{\xi^2}{[\lambda_\mu(\f{\pi}{2},\f{1}{\xi})]^2} =
\f{1}{[\lambda_{\tilde\mu}(\f{\pi}{2},\xi)]^2}\,.   
\end{equation}
Putting these results together one finds that
\begin{equation}
  \label{eq:app_LR_coeff}
  \begin{aligned}
    \Delta {\cal Z}_{\mu\nu}^{(LR)}(\xi) &=  \Delta {\cal
      Z}_{\mu\nu}(\textstyle\f{\pi}{2},\f{1}{\xi}) = \Delta {\cal
      Z}_{\tilde\mu\tilde\nu}(\textstyle\f{\pi}{2},\xi)\,,  \\
    \Delta {\cal G}_{\mu\nu}^{(LR)}(\xi) &=  \Delta {\cal
      G}_{\mu\nu}(\textstyle\f{\pi}{2},\f{1}{\xi}) +
    \displaystyle\f{1}{(4\pi)^2}\log\xi^2 = 
    \Delta {\cal G}_{\tilde\mu\tilde\nu}^{fin}(\textstyle\f{\pi}{2},\xi)\,.
  \end{aligned}
\end{equation}
Explicitly,
\begin{equation}
  \label{eq:largeT5_xi_0}
  \begin{aligned}
    \Delta {\cal G}_{23}^{(LR)}(\xi)
 & =    \Delta {\cal G}_{\perp\perp}^{(LR)}(\xi) = N_c\left[
\frac{11}{3}\tilde g_{00}({\textstyle\f{\pi}{2}})
-\frac{10}{3}\tilde g_{10}({\textstyle\f{\pi}{2}})
+\frac{2}{3}   \tilde g_{11}({\textstyle\f{\pi}{2}})
\right]+ o(\xi^0)\,,\\
\Delta {\cal G}_{42}^{(LR)}(\xi) & = \Delta {\cal G}_{43}^{(LR)}(\xi)=
\Delta {\cal G}_{12}^{(LR)}(\xi) = \Delta {\cal G}_{13}^{(LR)}(\xi)
= \Delta {\cal G}_{\parallel\perp}^{(LR)}(\xi) \\ &=
 N_c\left[
\frac{11}{3}\tilde g_{00}({\textstyle\f{\pi}{2}})
-\frac{5}{3}\tilde g_{10}({\textstyle\f{\pi}{2}})
\right] + o(\xi^0)\,,\\
\Delta {\cal G}_{41}^{(LR)}(\xi)
 & = \Delta {\cal G}_{\parallel\parallel}^{(LR)}(\xi) = o(\xi^0)\,,
  \end{aligned}
\end{equation}
for the contributions $\Delta{\cal G}_{\mu\nu}^{(LR)}$, and
\begin{equation}
  \label{eq:largeT5_xi_1}
  \begin{aligned}
\Delta {\cal Z}_{23}^{(LR)}(\xi) &= \Delta {\cal
  Z}_{\perp\perp}^{(LR),div}(\xi) + \Delta {\cal 
  Z}_{\perp\perp}^{(LR),fin}(\xi)  \,,  &\\ 
\Delta {\cal Z}_{42}^{(LR)}(\xi) &= \Delta {\cal Z}_{43}^{(LR)}(\xi) =
\Delta {\cal Z}_{12}^{(LR)}(\xi) = \Delta {\cal Z}_{13}^{(LR)}(\xi) \\
&=
\Delta {\cal Z}_{\parallel\perp}^{(LR),div}(\xi) + \Delta {\cal
  Z}_{\parallel\perp}^{(LR),fin}(\xi)  
\,,
\\
\Delta {\cal Z}_{41}^{(LR)}(\xi) & = \Delta {\cal
  Z}_{\parallel\parallel}^{(LR)}(\xi) =  o(\xi^0)\,,    
  \end{aligned}
\end{equation}
for the contributions $\Delta{\cal Z}_{\mu\nu}^{(LR)}$,
with divergent parts
\begin{equation}
  \label{eq:largeT5_xi_2}
  \begin{aligned}
\Delta {\cal Z}_{\perp\perp}^{(LR),div}(\xi) &=  \frac{N_c^2-1}{2N_c}
  \f{1}{2\pi}\log\xi^2 \,,  &\\
\Delta {\cal Z}_{\parallel\perp}^{(LR),div}(\xi) &= 
\f{N_c}{4}
\f{1}{4\pi}\log\xi^2\,,     
  \end{aligned}
\end{equation}
and finite parts
\begin{equation}
  \label{eq:largeT5_xi}
  \begin{aligned}
\Delta {\cal Z}_{\perp\perp}^{(LR),fin}(\xi) &=
\frac{N_c^2-1}{N_c}\big( z_{00} + 
  \tilde z_{10}({\textstyle\f{\pi}{2}})\big)
+ \f{N_c}{2}\big(
\tilde z_{00} -  \tilde 
  z_{10}({\textstyle\f{\pi}{2}})\big) + o(\xi^0)\,,  \\
\Delta {\cal Z}_{\parallel\perp}^{(LR),fin}(\xi) &= 
\f{N_c}{4}\big(
 z_{00}+ \tilde z_{00}({\textstyle\f{\pi}{2}})\big) +
\frac{N_c^2-2}{4N_c}z_{10}
+ o(\xi^0)\,.
\end{aligned}
\end{equation}
We also report the values of the ${\cal Z}$-integrals,
  \begin{equation}
    \label{eq:Zlargexi}
    \begin{aligned}
      {\cal Z}^{(LR)}_\parallel(\xi)&={\cal Z}^{(LR)}_4(\xi)=
      {\cal Z}^{(LR)}_1(\xi) = z_{10} + o(\xi^0)\,,\\
      {\cal Z}^{(LR)}_\perp(\xi) &=
      {\cal Z}^{(LR)}_2(\xi) =
      {\cal Z}^{(LR)}_3(\xi) = \f{1}{4\pi}\log\xi^2 + z_{00} + \tilde
      z_{10}(\textstyle\f{\pi}{2}) + o(\xi^0)\,.
    \end{aligned}
  \end{equation}

\secspace
\section{Abelian case}
\label{app:abelian}

In this Appendix we compute the Wilson-loop correlator considered in
Section \ref{sec:AC} in the compact $U(1)$ lattice theory and in the
weak-coupling limit.  
The starting point is the 4D anisotropic lattice formulation for the
$U(1)$ gauge group, 
\begin{equation}
  \label{eq:U1_action}
  S^{U(1)}_{\rm lat} = \f{1}{e^2}\sum_{n,\mu<\nu} C_{\mu\nu}
  \left(1 - \Re U_{\mu\nu}(n)\right)\,.
\end{equation}
Here the plaquettes are built with the $U(1)$
links $U_\mu(n) = \exp\{i\phi_\mu(n)\}$, and can be written as $\Re
U_{\mu\nu}(n)=\cos\Phi_{\mu\nu}(n)$, with $\Phi_{\mu\nu}(n)
=  \phi_\mu(n) + \phi_\nu(n+\hat\mu) - \phi_\mu(n+\hat\nu) -\phi_\nu(n)
$. The Haar measure is simply
$\int dU_\mu(n) = \int_{-\pi}^{+\pi}\f{d\phi_\mu(n)}{2\pi}$.
Setting
\begin{equation}
  \label{eq:tilt_U1_1}
  \begin{aligned}
 {\cal U}^{(1)}_{1}(n) &= 
U_4(n)U_1(n+\hat 4)=e^{i[\phi_4(n)+\phi_1(n+\hat 4)]} = e^{i\varphi^{(1)}_{1}(n)} \,, \\
    {\cal U}^{(1)}_{2}(n) &= 
    U_1(n)U_4(n+\hat 1)=e^{i[\phi_4(n+\hat 1)+\phi_1(n)]}= e^{i\varphi^{(1)}_{2}(n)}\,,\\
{\cal U}^{(2)}_{1}(n) &= 
U_4(n)U_1^*(n+\hat 4 -\hat 1)=e^{i[\phi_4(n)-\phi_1(n+\hat 4 -\hat
  1)]}= e^{i\varphi^{(2)}_{1}(n)}\,, \\
    {\cal U}^{(2)}_{2}(n) &= U_1^*(n-\hat 1)U_4(n-\hat 1)
=e^{i[\phi_4(n-\hat 1)-\phi_1(n-\hat 1)]}= e^{i\varphi^{(2)}_{2}(n)} \,,
  \end{aligned}
\end{equation}
the analogue of Eq.~\eqref{eq:tilt_link_unit} is here
\begin{equation}
  \label{eq:tilt_U1_3}
{\cal U}^{(j)}(n) =
\f{e^{i\varphi^{(j)}_{1}(n)}+e^{i\varphi^{(j)}_{2}(n)}}{|e^{i\varphi^{(j)}_{1}(n)}+
e^{i\varphi^{(j)}_{2}(n)}|}=
\exp\left(\f{i}{2}\big({\varphi^{(j)}_{1}(n)+\varphi^{(j)}_{2}(n)}\big)\right)
{\rm sign}\left(\cos \f{\Phi^{(j)}(n)}{2}\right)\,,
\end{equation}
where $\Phi^{(1)}(n)=\Phi_{41}(n)$ and $\Phi^{(2)}(n)=\Phi_{41}(n-\hat
1)$. The Wilson loops are written as $\W_{Lk} = e^{i\Omega_k}\sigma_k
T_k$, $k=1,2$, with  
\begin{equation}
  \label{eq:WLU1phase}
      \Omega_k = \f{1}{2}\sum_{j=-t_0}^{t_0-1} 
\left(\varphi^{(k)}_1 (j v_k+d_{k+}) + \varphi^{(k)}_2 (j v_k+d_{k+})
- \varphi^{(k)}_1 (j v_k+d_{k-}) - \varphi^{(k)}_2 (j v_k+d_{k-})
\right)\,,
\end{equation}
$\sigma_k$ the product of the sign factors appearing in
Eq.~\eqref{eq:tilt_U1_3}, and $T_k$ the contribution from the
transverse links. 

The calculation is greatly simplified if we take the limit
$\barT\to\infty$ first.\footnote{Since there is actually no continuum
  limit to be taken, in this case the complications of the non-Abelian
  case are absent.} Discarding the longitudinal-longitudinal plaquette
term, enforcing the triviality of the transverse links, and using $1 -
\Re U_{\mu\alpha} = \f{1}{2}|\Delta_\alpha^+U_\mu|^2$ for trivial $U_\alpha$
links, one ends up with 
\begin{equation}
  \label{eq:2D_action_U1}
 S^{U(1)}_{\rm lat} \mathop\to_{\barT\to\infty} \f{1}{2e^2}\sum_{\mu=4,1}
c^{(\mu)}
\sum_{n_\parallel,n_\perp}
\sum_{\alpha=2,3}
 |\Delta^+_\alpha U_\mu(n)|^2
\,, \quad c^{(4)}=\tan\f{\theta}{2}\,,~~
c^{(1)}=\cot\f{\theta}{2}\,,
\end{equation}
where $n_\parallel=(n_4,n_1)$ and $n_\perp = (n_2,n_3)$. The Wilson
loops simplify to $\W_{Lk} \to e^{i\Omega_k}\sigma_k$. 
Since there is no interaction between link variables living at
different sites of the longitudinal plane, and between $U_4$ and $U_1$
variables, one easily sees that the  ``tilted links'' of
Eq.~\eqref{eq:tilt_U1_3} 
interact with each other only if they are separated by at most one
lattice spacing in the longitudinal plane, which leads to
factorisation of the Wilson-loop correlation function and expectation
values.

It is convenient now to rescale the phases as
$\phi_\mu(n)=e\bar\phi_\mu(n_\parallel,x)$ with $x=e n_\perp$ (notice
that $x$ is dimensionless), in order to take the weak-coupling
limit. One then obtains for the action
\begin{equation}
  \label{eq:2D_action_U1_2}
 S^{U(1)}_{\rm lat}  \mathop \to_{\barT\to\infty,\,e\to 0} 
\sum_{\mu=4,1}
 c^{(\mu)} \sum_{n_\parallel}\int d^2x\sum_{\alpha=2,3}
\f{1}{2}[\de_\alpha \bar\phi_\mu(n_\parallel;x)]^2 
\,,
\end{equation}
and the integration measure in the weak-coupling limit becomes
\begin{equation}
  \label{eq:2D_action_U1_meas}
\int_{-\pi}^{+\pi}\f{d\phi_\mu(n)}{2\pi} \longrightarrow
\int_{-\infty}^{+\infty}   d\bar\phi_\mu(n_\parallel;x)\,,
\end{equation}
where we have omitted a factor $e/(2\pi)$ since it gets cancelled in
expectation values. We passed to the continuum notation for
simplicity: as the action is quadratic, the resulting continuum
Gaussian integrals are fully under control. The propagator is readily
obtained,  
\begin{equation}
  \label{eq:propU1full}
  D_{\mu\nu}(n_\parallel,m_\parallel;x,y) \equiv
 \la \bar\phi_\mu(n_\parallel;x) \bar\phi_\nu(m_\parallel;y)\ra
=
  \delta_{\mu\nu}\delta_{n_\parallel m_\parallel}
  \f{1}{c^{(\mu)}}D(x-y)\,,
\end{equation}
where $D(x)$ is the 2D scalar propagator,
\begin{equation}
  \label{eq:prop2D}
  D(x) =  -\f{1}{2\pi}\log |x|\,.
\end{equation}
From here on angular brackets without subscripts denote the
expectation value with respect to the action
Eq.~\eqref{eq:2D_action_U1_2}.  
In the weak-coupling limit, $\cos\Phi_{\mu\nu} = 1 +\Oc(e^2)$ and we
can neglect the sign factors in the expression for the Wilson loops,
i.e., $\W_{Lk} \to e^{i\Omega_k}$.  
Since the action is quadratic, one has for the relevant correlation
function as $e\to 0$
\begin{equation}
  \label{eq:corrfunc_wcl}
\lim_{\barT\to \infty} \f{  \la \W_1 \W_2
  \ra_{\theta,\bar T}}{  \la \W_1 \ra_{\theta,\bar T}\la\W_2
  \ra_{\theta,\bar T}} =
\f{e^{-\f{1}{2}\la (\Omega_1+\Omega_2)^2\ra}}{
e^{-\f{1}{2}\la\Omega_1^2\ra}
e^{-\f{1}{2}\la \Omega_2^2\ra}
} = e^{-\la \Omega_1\Omega_2\ra}\,.
\end{equation}
Using now the explicit expression for $\Omega_k$, see
Eqs.~\eqref{eq:tilt_U1_1} and \eqref{eq:WLU1phase}, and exploiting
the fact that the propagator is diagonal in the link directions and in
the longitudinal coordinates, a straightforward calculation gives
\begin{equation}
  \label{eq:corrfunc_wcl_2}
 \la \Omega_1\Omega_2\ra = \f{e^2}{2\pi} \cot\theta
\log\f{\Big|\vec z_\perp +\f{\vec R_{1\perp}}{2}+ \f{\vec R_{2\perp}}{2}\Big|
  \Big|\vec z_\perp -\f{\vec R_{1\perp}}{2}- \f{\vec R_{2\perp}}{2}\Big|}{
  \Big|\vec z_\perp +\f{\vec R_{1\perp}}{2}- \f{\vec R_{2\perp}}{2}\Big|
  \Big|\vec z_\perp -\f{\vec R_{1\perp}}{2}+ \f{\vec R_{2\perp}}{2}\Big|}\,,
\end{equation}
which agrees with the known result for ${\cal C}_E$ in the 4D $U(1)$
pure gauge theory in the continuum limit~\cite{Meggiolaro05}. Here we
have set $f_1=f_2=\f{1}{2}$ for convenience, without any loss of
information~\cite{GM2010}.

\secspace
\section{Large-$\barT$ limit of the $(\theta,\barT)$-dependent 
  action}  
\label{app:largeTaction}

For completeness, in this Appendix we report on the large-$\barT$
limit of the anisotropic action with anisotropy parameters
Eq.~\eqref{eq:anis_param}, discussed in Section \ref{sec:AC}. The idea
is that there could be some useful simplification if one takes $\barT
\to \infty$, corresponding to the limit of loops of infinite length,
before taking the continuum limit. In full
analogy with the discussion of Section \ref{sec:2dmodel}, in this
limit the action can be recast as that of a set of interacting
principal chiral models, which however live now in the transverse
plane at every site of the longitudinal plane. This is natural since
the limit $\barT\to\infty$ corresponds to taking the continuum limit
in the transverse plane at fixed longitudinal spacing $a_\parallel
\equiv \barT a$, i.e., the same situation of Section \ref{sec:2dmodel}
but reversing the roles of the longitudinal and the transverse planes.
In the large-$\barT$ limit, the longitudinal-transverse couplings can be
rewritten as follows, 
\begin{equation}
  \label{eq:largeT_beta4}
  \begin{aligned}
\beta_{2D}^{(4)}(a_\perp,a_\parallel,\theta) &=
\f{\beta_{42}}{2N_c}C_{42} = \f{\beta_{43}}{2N_c}C_{43} 
= \f{1}{2}\f{N_c}{4\pi}\log\f{1}{a_\perp\Lambda_{2D}^{(4)}(a_\parallel,\theta)}\,,
\\ \Lambda_{2D}^{(4)}(a_\parallel,\theta)&=\Lambda c
e^{-\f{8\pi}{N_c}\tan\f{\theta}{2}(\Delta {\cal G}_{4\perp}^{fin}(\infty) + 
    \Delta {\cal Z}_{4\perp}^{fin}(\infty))}
(a_\parallel\Lambda c)^{\f{16\pi\beta_0}{N_c}\tan\f{\theta}{2}-1}
\,, 
\\
\beta_{2D}^{(1)}(a_\perp,a_\parallel,\theta) &=
\f{\beta_{12}}{2N_c}C_{12} = \f{\beta_{13}}{2N_c}C_{13} 
= \f{1}{2}\f{N_c}{4\pi}\log\f{1}{a_\perp\Lambda_{2D}^{(1)}(a_\parallel,\theta)}\,,
\\ \Lambda_{2D}^{(1)}(a_\parallel,\theta)&=\Lambda c
e^{-\f{8\pi}{N_c}\cot\f{\theta}{2}(\Delta {\cal G}_{1\perp}^{fin}(\infty) + 
    \Delta {\cal Z}_{1\perp}^{fin}(\infty))}
(a_\parallel\Lambda c)^{\f{16\pi\beta_0}{N_c}\cot\f{\theta}{2}-1}\,,
  \end{aligned}
\end{equation}
which is precisely the form of the bare coupling as a function of the
lattice cutoff $a_\perp=a$ in the two-dimensional $SU(N_c)$ principal chiral
model, to one-loop accuracy (see, e.g., Ref.~\cite{Polyakov}). Here we
have neglected $o(T^0)$ terms. The
remaining couplings read, in the same limit and in the same approximation,
\begin{equation}
  \label{eq:largeT_others}
  \begin{aligned}
    \f{\beta_{41}}{2N_c}C_{41} &= \f{1}{\barT^2\sin\theta}
    \f{N_c^2-1}{2N_c}\f{1}{\pi\sin\theta}
    \log\f{1}{a_\perp\tilde\Lambda_{2D}(a_\parallel,\theta)} \equiv
    \f{1}{\barT^2}\tilde\beta_{2D}(a_\perp,a_\parallel,\theta) \,,\\   
    \tilde\Lambda_{2D}(a_\parallel,\theta) &=
\Lambda c (a_\parallel\Lambda c)^{\f{4\pi\beta_0
    N_c\sin\theta}{N_c^2-1}-1}
e^{-\f{2\pi
    N_c\sin\theta}{N_c^2-1}(\Delta {\cal G}_{41}^{fin}(\infty) +  
  \Delta {\cal Z}_{41 }^{fin}(\infty))}\,,\\
\f{\beta_{23}}{2N_c}C_{23} &=
\barT^2\sin\theta \beta_0\log\f{1}{(a_\parallel\Lambda c)^{2}} \equiv
\barT^2\hat\beta_{2D}(a_\parallel,\theta)\,.
  \end{aligned} 
\end{equation}
The action can be recast as follows,
\begin{equation}
  \label{eq:2D_theta_1loop_0}
S^{(2D)} = \sum_{n_\parallel}
S_\chi^{(4)}(n_\parallel)+S_\chi^{(1)}(n_\parallel) + S_{\rm int1}(n_\parallel) +
S_{\rm int2}(n_\parallel) \,,
\end{equation}
where $S_\chi^{(\mu)}$ correspond to principal chiral models,
\begin{equation}
  \label{eq:2D_theta_1loop_1}
S_\chi^{(\mu)}(n_\parallel) =
\beta_{2D}^{(\mu)}(a_\parallel,a_\perp,\theta)\sum_{n_\perp}\sum_{\alpha=2,3}
\tr\{[\Delta^+_\alpha 
U_\mu(n)][\Delta^+_\alpha U_\mu(n)]^\dag\}\,,
\end{equation}
and the mutual interactions are given by the remaining terms,
\begin{equation}
  \label{eq:2D_theta_1loop}
  \begin{aligned} 
S_{\rm int1}(n_\parallel) &=
\tilde\beta_{2D}(a_\parallel,a_\perp,\theta)\sum_{n_\perp}\f{a_\perp^2}{a_\parallel^2}
{\cal P}_{41}(n)\,,\\
S_{\rm int2}(n_\parallel) &=
\sum_{\mu=4,1} \beta_{2D}^{(\mu)}(a_\parallel,a_\perp,\theta)
\sum_{n_\perp}\sum_{\alpha=2,3} \left[2N_c {\cal
    P}_{\mu\alpha}(n) - \tr\{[\Delta^+_\alpha
U_\mu(n)][\Delta^+_\alpha U_\mu(n)]^\dag\}
\right] \\ & \phantom{=}
+\hat\beta_{2D}(a_\perp,\theta)
\sum_{n_\perp}2N_c\f{a_\parallel^2}{a_\perp^2}{\cal
  P}_{23}(n)\,. 
  \end{aligned}
\end{equation}
The two-dimensional scales $\Lambda_{2D}^{(4,1)}$ and
$\tilde\Lambda_{2D}$ have prescribed
values that depend on $\Lambda$, which is set in the 4D theory, and on
$a_\parallel$, which has to be taken to zero at the end of the
calculation. However, the average plaquette terms to lowest
order and for large $\barT$ read [see Eqs.~\eqref{eq:plaq_2} and
\eqref{eq:largeT3}] 
\begin{equation}
  \label{eq:plaq_Ttheta}
  \begin{aligned}
    \la {\cal P}_{41}\ra    &\simeq g^2\f{N_c^2-1}{N_c}
    \f{\log\barT^2}{4\pi\sin\theta}\,, &&&
    \la {\cal P}_{23}\ra    &\simeq
    g^2\f{N_c^2-1}{N_c}\f{z_{10}}{\barT^2\sin\theta}  \,,\\
    \la {\cal P}_{4\perp}\ra &\simeq
    g^2\f{N_c^2-1}{2N_c}\cot\f{\theta}{2}z_{10}\,,&&&
    \la {\cal P}_{1\perp}\ra &\simeq
    g^2\f{N_c^2-1}{2N_c}\tan\f{\theta}{2}z_{10}\,, 
  \end{aligned}
\end{equation}
so that the range of applicability of perturbation theory is limited
by $g^2\log\barT\ll 1$; more precisely, besides $a_\parallel \gg
a_\perp$ one needs $a_\parallel\Lambda\ll (a_\perp\Lambda)^{1-\gamma}$
for some $\theta$-dependent $\gamma$, which prevents from taking the
continuum limit in the transverse plane independently from the
longitudinal plane.


\newpage


\begin{thebibliography}{99}

\bibitem{MP}
  C.~J.~Morningstar and M.~J.~Peardon,
  Phys.\ Rev.\ D {\bf 60} (1999) 034509
  [hep-lat/9901004].

\bibitem{Lin:2008pr}
H.~W.~Lin, S.~D.~Cohen, J.~Dudek, R.~G.~Edwards, B.~Jo\'o,
D.~G.~Richards, J.~Bulava, J.~Foley, C.~Morningstar, E.~Engelson,
S.~Wallace, K.~J.~Juge, N.~Mathur, M.~J.~Peardon and S.~M.~Ryan
[Hadron Spectrum Collaboration],
  Phys.\ Rev.\ D {\bf 79} (2009) 034502
  [arXiv:0810.3588 [hep-lat]].

\bibitem{Namekawa:2001ih}
Y.~Namekawa, S.~Aoki, R.~Burkhalter, S.~Ejiri, M.~Fukugita, S.~Hashimoto,
 N.~Ishizuka, Y.~Iwasaki, K.~Kanaya, T.~Kaneko, Y.~Kuramashi, V.~Lesk,
 M.~Okamoto, M.~Okawa, Y.~Taniguchi, A.~Ukawa, and T.~Yoshie [CP-PACS
 Collaboration],  
  Phys.\ Rev.\ D {\bf 64} (2001) 074507
  [hep-lat/0105012].

\bibitem{Borsanyi:2014vka}
  S.~Bors\'anyi, S.~D\"urr, Z.~Fodor, C.~Hoelbling, S.~D.~Katz, S.~Krieg,
  S.~Mages, D.~N\'ogr\'adi, 
  A.~P\'asztor, A.~ Sch\"afer, K.~K.~Szab\'o,
  B.~C.~T\'oth, and N.~Trombit\'as,
  JHEP {\bf 1404} (2014) 132
  [arXiv:1401.5940 [hep-lat]].


\bibitem{Bur} G.~Burgio, A.~Feo, M.~J.~Peardon and S.~M.~Ryan,
  Phys.\ Rev.\ D {\bf 67} (2003) 114502 [hep-lat/0303005].

\bibitem{Verlinde}
H.~Verlinde and E.~Verlinde, hep-th/9302104.

\bibitem{Arefeva1} I.~Ya.~Aref'eva, Phys.\ Lett.\ B {\bf 325} (1994)
  171 [hep-th/9311115].

\bibitem{Arefeva2} I.~Ya.~Aref'eva, Phys.\ Lett.\ B {\bf 328} (1994) 411
  [hep-th/9306014].

\bibitem{Orland:2008vg}
  P.~Orland,
  Phys.\ Rev.\ D {\bf 77} (2008) 056004
  [arXiv:0801.0389 [hep-ph]].

\bibitem{Orland} P.~Orland and J.~Xiao, Phys.\ Rev.\ D {\bf 80} (2009) 016005
  [arXiv:0901.2955]. 

\bibitem{Cubero:2011ut}
  A.~Cort\'es Cubero and P.~Orland,
  Phys.\ Rev.\ D {\bf 84} (2011) 065034
  [arXiv:1104.1168 [hep-th]].

\bibitem{Cubero:2012nw}
  A.~Cort\'es Cubero and P.~Orland,
  arXiv:1203.5141 [hep-th].


\bibitem{GM2009}
  M.~Giordano and E.~Meggiolaro, Phys.\ Lett.\ B {\bf 675} (2009) 123
  [arXiv:0902.4145 [hep-ph]]. 

  
\bibitem{analytic1}
  E.~Meggiolaro, Z.\ Phys.\ C {\bf 76} (1997) 523 [hep-th/9602104].

\bibitem{analytic2}
  E.~Meggiolaro, Eur.\ Phys.\ J.\ C {\bf 4} (1998) 101 [hep-th/9702186].

\bibitem{analytic3}
  E.~Meggiolaro, Nucl.\ Phys.\ B {\bf 625} (2002) 312 [hep-ph/0110069].

\bibitem{Meggiolaro05}
  E.~Meggiolaro, Nucl.\ Phys.\ B {\bf 707} (2005) 199 [hep-ph/0407084].

\bibitem{crossing1}
  M.~Giordano and E.~Meggiolaro, Phys.\ Rev.\ D {\bf 74} (2006) 016003
  [hep-ph/0602143].

\bibitem{crossing2}
  E.~Meggiolaro, Phys.\ Lett.\ B {\bf 651} (2007) 177 [hep-ph/0612307]. 

\bibitem{Nachtmann91}
  O.~Nachtmann, Ann.\ Phys.\ (N.Y.) {\bf 209} (1991) 436.

\bibitem{DFK}
  H.~G.~Dosch, E.~Ferreira and A.~Kr{\"a}mer, Phys.\ Rev.\ D {\bf 50}
  (1994) 1992 [hep-ph/9405237].

\bibitem{Nachtmann97}
  O.~Nachtmann, Lect.\ Notes Phys.\ {\bf 496} (1997) 1
[hep-ph/9609365].

\bibitem{BN}
  E.~R.~Berger and O.~Nachtmann, Eur.\ Phys.\ J.\ C {\bf 7} (1999) 459
  [hep-ph/9808320]. 

\bibitem{Dosch}
  H.~G.~Dosch, in {\it At the frontier of Particle Physics -- Handbook of QCD
    (Boris Ioffe Festschrift)}, edited by M.~Shifman (World
  Scientific, Singapore, 2001), vol.\ 2, 1195--1236.

\bibitem{LLCM1}
  A.~I.~Shoshi, F.~D.~Steffen and H.~J.~Pirner, Nucl.\ Phys.\ A {\bf 709}
  (2002) 131 [hep-ph/0202012]. 

\bibitem{pomeron-book}
  S.~Donnachie, G.~Dosch, P.~Landshoff and O.~Nachtmann, {\it Pomeron Physics
    and QCD} (Cambridge University Press, Cambridge, England, 2002).

\bibitem{reggeon} M.~Giordano,
  JHEP {\bf 07} (2012) 109
   [Erratum-ibid.\ {\bf 1301} (2013) 021]  
   [arXiv:1204.3772 [hep-ph]].


\bibitem{sigtot}  M.~Giordano and E.~Meggiolaro, 
JHEP {\bf 03} (2014)  002 [arXiv:1311.3133 [hep-ph]]. 

\bibitem{DeW1} B.~S.~DeWitt, Phys.\ Rev.\ {\bf 162} (1967) 1195.

\bibitem{DeW2} B.~S.~DeWitt, Phys.\ Rev.\ {\bf 162} (1967) 1239.

\bibitem{Abbott1}
  L.~F.~Abbott,
  Nucl.\ Phys.\ B {\bf 185} (1981) 189.


\bibitem{Abbott2}
  L.~F.~Abbott, M.~T.~Grisaru and R.~K.~Schaefer,
  Nucl.\ Phys.\ B {\bf 229} (1983) 372.


\bibitem{DG} R.~Dashen and D.~J.~Gross, Phys.\ Rev.\ D {\bf 23} (1981) 2340.

\bibitem{HH} A.~Hasenfratz and P.~Hasenfratz, Nucl.\ Phys.\ B {\bf 193}
  (1981) 210. 

\bibitem{Karsch} F.~Karsch, Nucl.\ Phys.\ B {\bf 205} (1982) 285. 

\bibitem{GAKA} A.~Gonzalez-Arroyo and C.~P.~Korthals Altes, Nucl.\ Phys.\ B
  {\bf 205} (1982) 46.

\bibitem{LW}
  M.~L\"uscher and P.~Weisz,
  Nucl.\ Phys.\ B {\bf 452} (1995) 213
  [hep-lat/9504006].

\bibitem{LWml}
  M.~L\"uscher and P.~Weisz,
  JHEP {\bf 09} (2001) 010  [hep-lat/0108014].

\bibitem{BGF-rev} L.~F.~Abbott, Acta Phys.\ Polon.\ B {\bf 13} (1982) 33.

\bibitem{Z-J} J.~Zinn-Justin, {\it Quantum f\mbox{}ield theory and critical
  phenomena} (Oxford University Press, New York, 2002). 

\bibitem{tH} G.~'t Hooft, unpublished. See \emph{Proceedings of the
    Colloquium on Renormalization of Yang--Mills Fields and
    Applications to Particle Physics}, edited by C.~P.~Korthals Altes
  (Marseille, 1972). 

\bibitem{Gr} D.~J.~Gross and F.~Wilczek, {Phys. Rev. Lett.} {\bf 30}
  (1973) 1343. 

\bibitem{Po} H.~D.~Politzer, {Phys. Rev. Lett.} {\bf 30} (1973) 1346.

\bibitem{GPvB} M.~Garc\'ia P\'erez and P.~van Baal, Phys. Lett. B {\bf
  392} (1997) 163 [hep-lat/9610036].


\bibitem{DHHS} I.~T.~Drummond, A.~Hart, R.~R.~Horgan and L.~C.~Storoni,
  Phys. Rev. D {\bf 66} (2002) 094509 [hep-lat/0208010].

\bibitem{RD}
  M.~Rueter and H.~G.~Dosch,
  Phys.\ Lett.\ B {\bf 380} (1996) 177
  [hep-ph/9603214].


\bibitem{LLCM2}
  A.~I.~Shoshi, F.~D.~Steffen, H.~G.~Dosch and H.~J.~Pirner,
  Phys. Rev. D {\bf 68} 
  (2003) 074004 [hep-ph/0211287].

\bibitem{ILM}
  E.~Shuryak and I.~Zahed, Phys. Rev. D {\bf 62} (2000) 085014
  [hep-ph/0005152]. 

\bibitem{JP}
  R.~A.~Janik and R.~Peschanski, Nucl. Phys. B {\bf 565} (2000) 193
  [hep-th/9907177]. 

\bibitem{JP2}
  R.~A.~Janik and R.~Peschanski, Nucl. Phys. B {\bf 586} (2000) 163
  [hep-th/0003059].

\bibitem{Janik}
  R.~A.~Janik, Phys. Lett. B {\bf 500} (2001) 118 [hep-th/0010069].

\bibitem{GP2010}
  M.~Giordano and R.~Peschanski, JHEP 
  {\bf 05} (2010) 037 
[arXiv:1003.2309 [hep-ph]].

\bibitem{GM2008}
  M.~Giordano and E.~Meggiolaro, Phys. Rev. D {\bf 78} (2008) 074510
  [arXiv:0808.1022 [hep-lat]]. 

\bibitem{GM2010}
  M.~Giordano and E.~Meggiolaro, Phys. Rev. D {\bf 81} (2010) 074022
  [arXiv:0910.4505 [hep-ph]]. 

\bibitem{GMM} M.~Giordano, E.~Meggiolaro and N.~Moretti, JHEP 
  {\bf 09} (2012) 031 [arXiv:1203.0961 [hep-ph]].

\bibitem{sigtot_comments} M.~Giordano and E.~Meggiolaro,
  Phys. Lett. B {\bf 744} (2015) 263 
  [arXiv:1411.0553 [hep-ph]].

\bibitem{DV} V.~S.~Dotsenko and S.~N.~Vergeles, Nucl. Phys. B {\bf 169}
(1980) 527.

\bibitem{BNS} R.~A.~Brandt, F.~Neri and M.~A.~Sato, Phys. Rev. D {\bf 24}
(1981) 879.

\bibitem{Pot1} T.~Appelquist and W.~Fischler, Phys. Lett. B {\bf 77}
  (1978) 405.

\bibitem{Pot2} G.~Bhanot, W.~Fischler and S.~Rudaz, Nucl. Phys. B {\bf
    155} (1979) 208.

\bibitem{Pot3} M.~E.~Peskin, Nucl. Phys. B {\bf 156} (1979) 365.

\bibitem{Pot4} G.~Bhanot and M.~E.~Peskin, Nucl. Phys. B {\bf 156}
  (1979) 391.

\bibitem{uni-st1}   M.~Ishida and K.~Igi, Phys. Lett. B {\bf 670} (2009) 395
  [arXiv:0809.2424 [hep-ph]].

\bibitem{uni-st2}
  K.~A.~Olive {\it et al.}  [Particle Data Group Collaboration],
  Chin.\ Phys.\ C {\bf 38} (2014) 090001.

\bibitem{Tit} E.~C.~Titchmarsh, {\it The Theory of Functions}, 2nd
  edition (Oxford University Press, London, 1939).


 
\bibitem{Polyakov}  A.~M.~Polyakov, 
  {\it Gauge Fields and Strings} (Harwood Academy Publisher, Chur,
  Switzerland, 1987).




\bibitem{GradRhiz} I.~S.~Gradshteyn and I.~M.~Ryzhik, {\it Table of integrals,
    series and products} (Academic Press, New York, 1980).


\end{thebibliography}
\end{document}